\documentclass{emulateapj}
\newcommand{\ms}{m s$^{-1}$ }

\usepackage{graphicx}
\usepackage{color}
\usepackage{amsmath}
\usepackage{amssymb}
\usepackage{hyperref}
\usepackage{epstopdf}
\usepackage{upgreek}
\usepackage{xcolor}
\hypersetup{
    colorlinks,
    linkcolor={red!80!black},
    citecolor={blue!80!black},
    urlcolor={blue!80!black}
}
\shorttitle{A compact, efficient fiber double scrambler}
\shortauthors{Halverson \& Roy et al.}
\slugcomment{Accepted for publication in ApJ}

\begin{document}

\title{An efficient, compact, and versatile fiber double scrambler for high precision radial velocity instruments}

\author{
Samuel Halverson\altaffilmark{1,2,3,\dag}, 
Arpita Roy\altaffilmark{1,2,3,\dag},
Suvrath Mahadevan\altaffilmark{1,2,3}, 
Lawrence Ramsey\altaffilmark{1,2}, 
Eric Levi\altaffilmark{1},	\\	
Christian Schwab\altaffilmark{1,2,4}, 
Fred Hearty\altaffilmark{1,2},
Nick MacDonald\altaffilmark{5}
}

\altaffiltext{\dag}{These authors contributed equally to this work}
\altaffiltext{1}{Department of Astronomy \& Astrophysics, The Pennsylvania State University, 525 Davey Lab, University Park, PA 16802, USA; {shalverson@psu.edu}, {aur17@psu.edu}}
\altaffiltext{2}{Center for Exoplanets \& Habitable Worlds, University Park, PA 16802, USA}
\altaffiltext{3}{Penn State Astrobiology Research Center, University Park, PA 16802, USA}
\altaffiltext{4}{Sagan Fellow}
\altaffiltext{5}{Department of Astronomy, University of Washington, Seattle, WA 98195, USA}


\begin{abstract}
We present the design and test results of a compact optical fiber double-scrambler for high-resolution Doppler radial velocity instruments. This device consists of a single optic: a high-index $n$$\sim$2 ball lens that exchanges the near and far fields between two fibers.  When used in conjunction with octagonal fibers, this device yields very high scrambling gains and greatly desensitizes the fiber output from any input illumination variations, thereby stabilizing the instrument profile of the spectrograph and improving the Doppler measurement precision. The system is also highly insensitive to input pupil variations, isolating the spectrograph from telescope illumination variations and seeing changes. By selecting the appropriate glass and lens diameter the highest efficiency is achieved when the fibers are practically in contact with the lens surface, greatly simplifying the alignment process when compared to classical double-scrambler systems. This prototype double-scrambler has demonstrated significant performance gains over previous systems, achieving scrambling gains in excess of 10,000 with a throughput of $\sim$87\% using uncoated Polymicro octagonal fibers. Adding a circular fiber to the fiber train further increases the scrambling gain to $>$20,000, limited by laboratory measurement error. While this fiber system is designed for the Habitable-zone Planet Finder spectrograph, it is more generally applicable to other instruments in the visible and near-infrared. Given the simplicity and low cost, this fiber scrambler could also easily be multiplexed for large multi-object instruments.
\end{abstract}

\section{Introduction}
The field of radial velocity (RV) extrasolar planet detection is poised on the discovery of the first true Earth analog, having so far been denied by the exquisite levels of Doppler precision needed to detect such a planet. The expectedRV amplitude of an Earth-size planet orbiting in the habitable-zone of a Sun-like star is $\sim$10~cm~s$^{-1}$, far below the precision achievable with the current generation of dedicated Doppler spectrographs. Attaining this precision goal will require, among other things, an unprecedented level of long-term instrument stability. One of the major technical challenges in improving the stability of these instruments is stabilizing the instrument illumination. Conventional slit-fed spectrographs have instrument point-spread functions (PSF) that vary significantly with seeing conditions and telescope guiding variability, thus limiting their ultimate RV precision. Consequently, these instruments require specialized methods of simultaneous calibration and data reduction to retrospectively discern stellar RV signals from instrument illumination variations \citep{butler:1996}.

Optical fibers offer a convenient and efficient method of transporting light from telescopes, allowing instruments to be physically decoupled from the telescope focus. While the addition of a conventional circular fiber provides a significant improvement in instrument PSF stability, the output of a typical astronomical fiber is still sensitive to the flux distribution incident on the fiber from the telescope. As this input illumination varies (e.g. due to guiding errors, seeing, or telescope pupil variations), the output intensity distribution also varies. This is particularly an issue for fiber-fed Doppler RV instruments since any illumination variations in the fiber output directly manifest as changes in the instrument profile. This effect cannot be calibrated with standard emission wavelength references that use a dedicated calibration fiber, since the output illumination is specific to the fiber and source and thus the calibration source does not accurately trace any variations in the stellar illumination propagating through the object fiber. In fact, the problem of spectrograph illumination is currently acknowledged to be one of the precision-limiting factors in Doppler spectroscopy, along with scientific source stability and wavelength calibration \citep{Pepe:2008,Avila:2008,pepe:2014}.  

Step-index multimode fibers have long been used as optical image scramblers to decouple the distribution of flux in the output beam from the illumination pattern at the telescope focal plane \citep{Heacox:1986}. In agreement with theory \citep{Heacox:1987}, repeated laboratory tests show that standard circular fibers yield a high degree of azimuthal scrambling, but minimal radial scrambling \citep[e.g.][]{Avila:2006,Avila:2008}. In more recent years, fibers with non-circular cores have been tested to some degree and demonstrated significant improvement over standard circular core fibers \citep[e.g.][]{Avila:2012,Spronck:2012}.  However, a single non-circular core fiber cannot deliver the kind of homogeneity, in both near and far-field, required by spectrographs for high-precision RV measurements below the 1 \ms level. These instruments require high levels of scrambling in both the near and far fields of the beam, without significantly compromising overall throughput. 

The fiber near-field, defined as the positional intensity distribution across the fiber face, is re-imaged in the spectrograph focal plane while the far-field is the angular distribution of the propagating beam. Any variations in the near-field will manifest as PSF variations, degrading overall instrument stability and measurement precision. The far-field pattern is projected onto the grating, and time-varying inhomogeneities that change groove illumination can introduce variable wavelength shifts at the detector. Changes in the far field also affect the illumination of the camera and other spectrograph optics, and therefore can change the amount of abberations contributing to the PSF shape, causing velocity shifts of spectral lines on the detector. Any spurious profile changes in the spectra diminish RV precision, and hence stabilizing both the near-field and far-field of the fiber is key for precision Doppler spectroscopy. 

A popular optical method of increasing the scrambling gain (SG) of a system is to introduce an optical double scrambler into the instrument fiber delivery system that exchanges the near and far fields by means of a lens relay \citep{Connes:1985,Brown:1990, Hunter:1992}. Coupling octagonal fibers on either side of a double scrambler not only boosts the level of scrambling, but also stabilizes both the near and far field illumination patterns. The inclusion of even a single octagonal fiber with a double scrambler has been shown to significantly improve on-sky RV precision and instrument stability \citep{Bouchy:2013}. 

Here we present a compact, easy-to-align double scrambler system based on a high refractive index ball lens. This device, when combined with octagonal fibers, yields SGs in excess of 10,000. The measured efficiency of the scrambler assembly is 87\% in the near-infrared (NIR), close to the theoretical maximum of the system in its current state (with uncoated fiber faces). The addition of a circular fiber increases the SG of the system to $>$20,000. The prototype device used in the majority of testing uses all off-the-shelf components and takes advantage of existing high-tolerance components used in commercial telecommunications industries. Our prototype scrambler is customized for the NIR (0.82 - 1.3 $\upmu$m), but with careful selection of lens material, this design could easily be adapted to visible wavelengths (380 - 900 nm) given the recent availability of broadband, high-index glasses from a variety of manufacturers. The simple design and minimal alignment requirements, when compared to classical double scramblers, implies that this device could easily be replicated for multiplexed fiber instruments on larger scales.

\section{Scrambling on the Hobby-Eberly Telescope}

Our design of a high scrambling gain system is motivated by the requirements of the Habitable Zone Planet Finder \citep[HPF,][]{Mahadevan:2014}, a stabilized fiber-fed precision RV spectrograph operating in the NIR between 0.82 to 1.3 $\upmu$m, currently being built for the 10-m Hobby-Eberly Telescope (HET). While scrambling is important for any high precision RV instrument, it is critical for an instrument on the HET since the fixed-elevation telescope design yields a highly variable entrance pupil relative to other telescopes. Stellar tracking is accomplished by moving a spherical aberration corrector (SAC) and prime focus instrument package (PFIP) over the primary spherical mirror. The actual telescope pupil is therefore defined by the geometric intersection of the SAC, wide-field corrector (WFC) pupil, and the HET primary, and both its orientation and shape change continuously during a track \citep{Lee:2010}. Without adequate scrambling, the changing pupil introduces large errors in the radial velocity measurements, and greatly undermines our primary science goals. During a set of observations, these pupil variations may average out over the track as a function of the target elevation and exposure time. Nevertheless, we aim for maximal scrambling gains such that the spectrograph PSF is desensitized to even the most extreme cases of input illumination change.

\section{Scrambling Differentiated from Modal Noise}
It should be noted that, in the context of this paper, \lq{}scrambling\rq{} differs from fiber speckle noise, also called \lq{}modal-noise\rq{}, which is due to the time-dependent variation of the speckle pattern produced by the finite number traversing modes within a multi-mode fiber interfering at the fiber exit boundary \citep{Baudrand:2001,McCoy:2012}. Scrambling here refers only to the decorrelation of fiber output illumination from input variations (or a re-arrangement of existing modes), and is a {\em static} correction that does not account for the time variation of modal power and phase distribution between exposures caused by the changing internal geometry of light rays. Proper modal noise mitigation requires a temporal redistribution of modes, commonly achieved through physical agitation of the fiber \citep{Baudrand:2001}. This forces the light to populate different modes (including unpopulated ones) with a large distribution of phases, and reduces modal noise by varying the number and phase distribution of modes per exposure. 

While poor scrambling changes the shape of the instrument profile and can shift its centroid, the effect of modal noise is to introduce systematic {\em uncertainty} in the instrument profile over time. Both are detrimental to high-precision RV measurements but are different phenomenon that do not have a single solution. We emphasize that unlike scrambling, modal noise mitigation requires a time variable correction that cannot be supplied by static octagonal fibers or the presence of a double scrambler alone. In addition to the scrambling system described here, HPF will have a modal noise agitator for its fibers \citep{McCoy:2012,Roy:2014}. We are also exploring new ways of suppressing modal noise for highly coherent calibration sources, where this effect is exacerbated \citep{Halverson:2014, Mahadevan:2014a}.

\section{Scrambling requirements for RV instruments}

The classical metric used to quantify scrambling performance is {\em scrambling gain} (SG). The formal definition of SG is the ratio of the relative displacement of the illumination distribution at the input of the fiber to the relative displacement of the illumination distribution at the output \citep{Avila:2008}:

\begin{equation}
\mathrm{SG} = \frac{{\Delta}d_\mathrm{input}/D_\mathrm{input}}{{\Delta}d_\mathrm{output}/D_\mathrm{output}},
\end{equation}

\noindent where ${\Delta}d_\mathrm{input}$ is the illumination shift on the fiber, $D_\mathrm{input}$ is the fiber diameter, ${\Delta}d_\mathrm{output}$ is the measured output centroid shift, and $D_\mathrm{output}$ is the width of the fiber (for near-field measurements) or the width of the output pupil (for far-field measurements).
Several factors influence the minimum scrambling gain required to reach a given velocity precision goal, including telescope guiding precision, spectrograph dispersion, and focal plane sampling. In the case of near-field scrambling (i.e. guiding induced errors) the expected velocity error is estimated in Equation~\ref{eq:scram_vel}:

\begin{equation}
{\sigma_{v,\mathrm{\ near-field}}} = \frac{c}{R_\mathrm{eff}}\frac{\Delta\theta}{\theta}\frac{1}{\mathrm{SG}},
\label{eq:scram_vel}
\end{equation}

\noindent where $R_\mathrm{eff}$ is the effective instrument resolution, $\Delta\theta$ is the telescope guiding error, $\theta$ is the size of the fiber on-sky, and SG is the scrambling gain of the fiber delivery system. Note that for instruments that rely on the use of a slit, $R_\mathrm{eff}$ is the effective resolution of the slit-less fiber in the detector focal plane: $R_\mathrm{eff} = R_\mathrm{slit} \times(w_\mathrm{slit}/w_\mathrm{fiber})$, where $R_\mathrm{slit}$ is the slit-limited instrument resolution, $w_\mathrm{slit}$ is the slit width and $w_\mathrm{fiber}$ is the fiber core diameter.

\begin{figure}
\begin{center}
\includegraphics[width=3.4in]{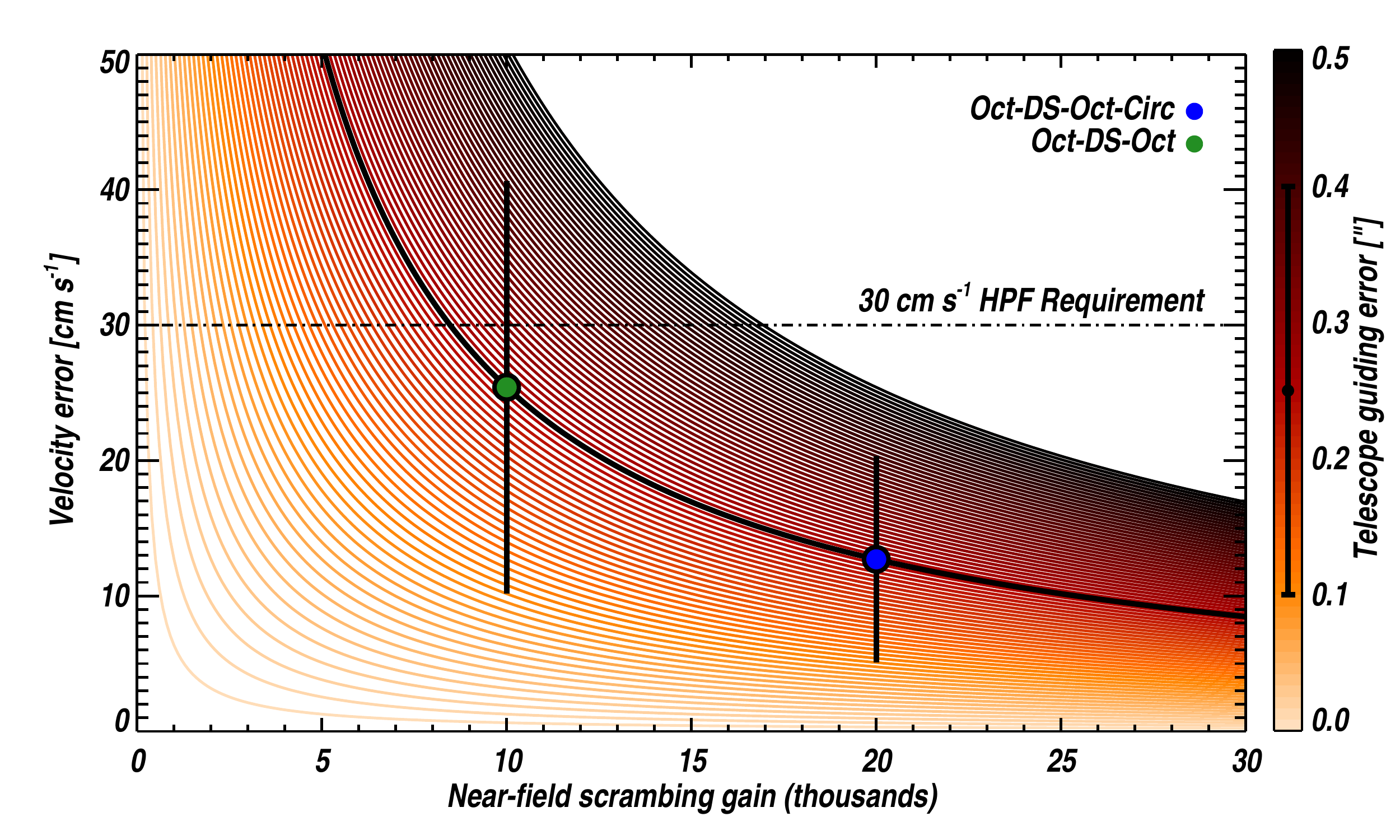}
\caption{Theoretical guiding-induced velocity error for HPF for a range of guiding precisions. Colored curves represent different guiding errors. The solid black line is the expected velocity error based on the HET guiding precision (0.25" RMS). The colored dots are theoretical velocity errors associated with the near-field scrambling measurements for fiber configurations measured in the laboratory (see Section~\ref{sec:results}). Vertical bars show an estimated range of possible guiding precisions for the HET during a given exposure.}
\label{fig:theo_scram}
\end{center}
\end{figure}

In the case of the HPF spectrograph on HET, with expected guiding precision of 0.25" \citep{lee:2012}, 1.7" fibers on sky (300 $\upmu$m core diameter), a 100 $\upmu$m slit, and an operating resolution of 50,000, a scrambling gain of $\sim$9,000 is required to achieve 30~cm~s$^{-1}$ stability. In practice, we aim for scrambling gains above 18,000 to ensure comparable velocity precisions even with guiding accuracies as low as 0.5'' (see Figure~\ref{fig:theo_scram}). This significantly reduces the risk of having guiding error be a dominant factor in the overall instrument velocity error budget. Since the primary science of HPF relies on velocity precision, it was incumbent upon us to meet these stringent, albeit previously undemonstrated, levels of scrambling gain.

\begin{figure}
\begin{center}
\includegraphics[width=2.2in]{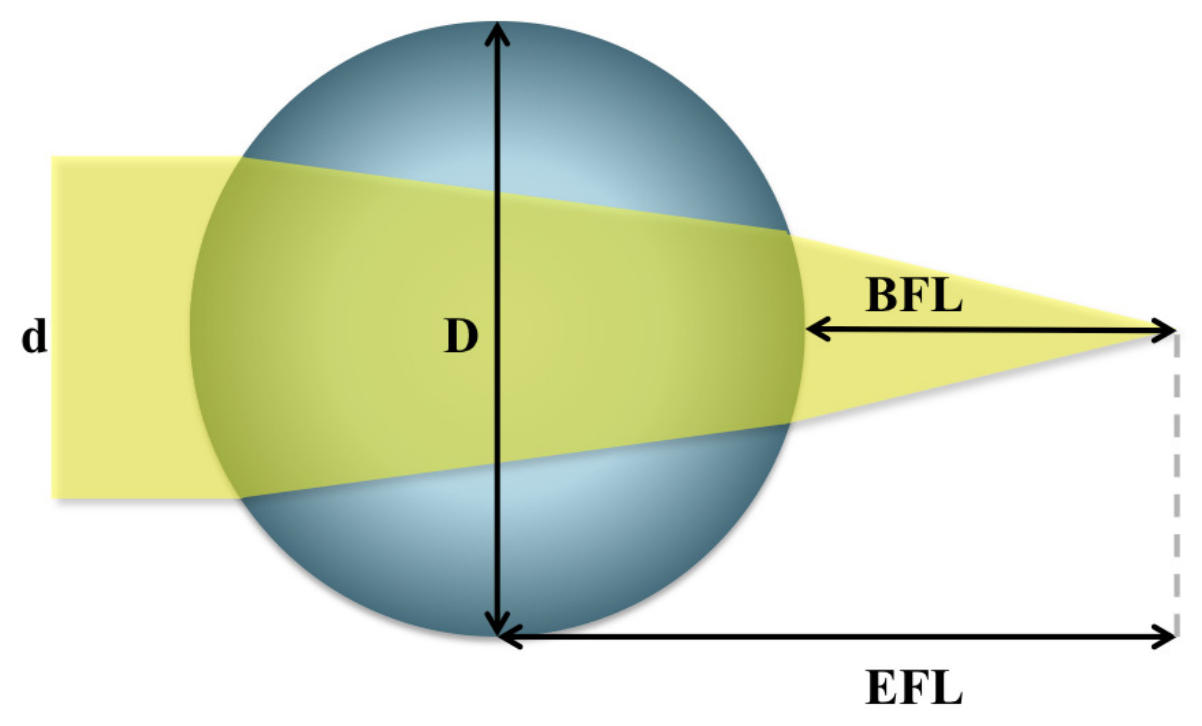}
\caption{Basic characteristics of a ball lens scrambler, including diameter of the lens ($D$), diameter of input fiber ($d$), effective focal length (EFL) and back focal length (BFL). Adapted from Edmund Optics technical resources.}
\label{fig:ball_cartoon}
\end{center}
\end{figure}

\section{Description of the Ball Lens Scrambler}
Ball lenses have previously been considered for astronomical scrambling devices, but have had varying success due to the unavailability of optimally sized spheres, alignment constraints, and high throughput losses \citep{Avila:2008,Barnes:2010,Spronck:2013}. Our design of the ball lens scrambler is motivated by the current availability of high-index glasses that provide shorter back focal lengths and simplify fiber coupling. A ball lens inherently either focuses or collimates light according to input source geometry, and in interchanging angle and position essentially acts as a single lens double scrambler. 

The back focal length of the lens (BFL, see Figure~\ref{fig:ball_cartoon}) is calculated from characteristic parameters of a ball lens using Equation~\ref{eq:bfl}, where $D$ is the diameter of the lens, $n$ is the refractive index of the glass used, and EFL is the effective focal length\footnote{http://www.edmundoptics.com/}.
\begin{equation}
\mathrm{BFL} =  \mathrm{EFL} - \frac{D}{2},\ \ \mathrm{EFL} =  \frac{nD}{4(n-1)}.
\label{eq:bfl}
\end{equation}
\begin{equation}
D =  \frac{2d(n-1)}{n}\sqrt{\frac{1}{\mathrm{NA}^2}-1}\ \approx \frac{2d(n-1)}{n \times \mathrm{NA}}
\label{eq:size}
\end{equation}
The remarkable providence in our ball scrambler design arises from the fact that when material of $n \sim 2$ is used, the BFL goes to {\em zero}, irrespective of ball diameter. This implies optimum coupling is achieved when the entrance and exit fibers are in direct contact with the lens surface,  simplifying the overall scrambler design. This greatly mitigates the need to position and control the ball lens at a specific distance, centered with high tolerance on the fibers.

\begin{figure}
\begin{center}
\includegraphics[width=3.4in]{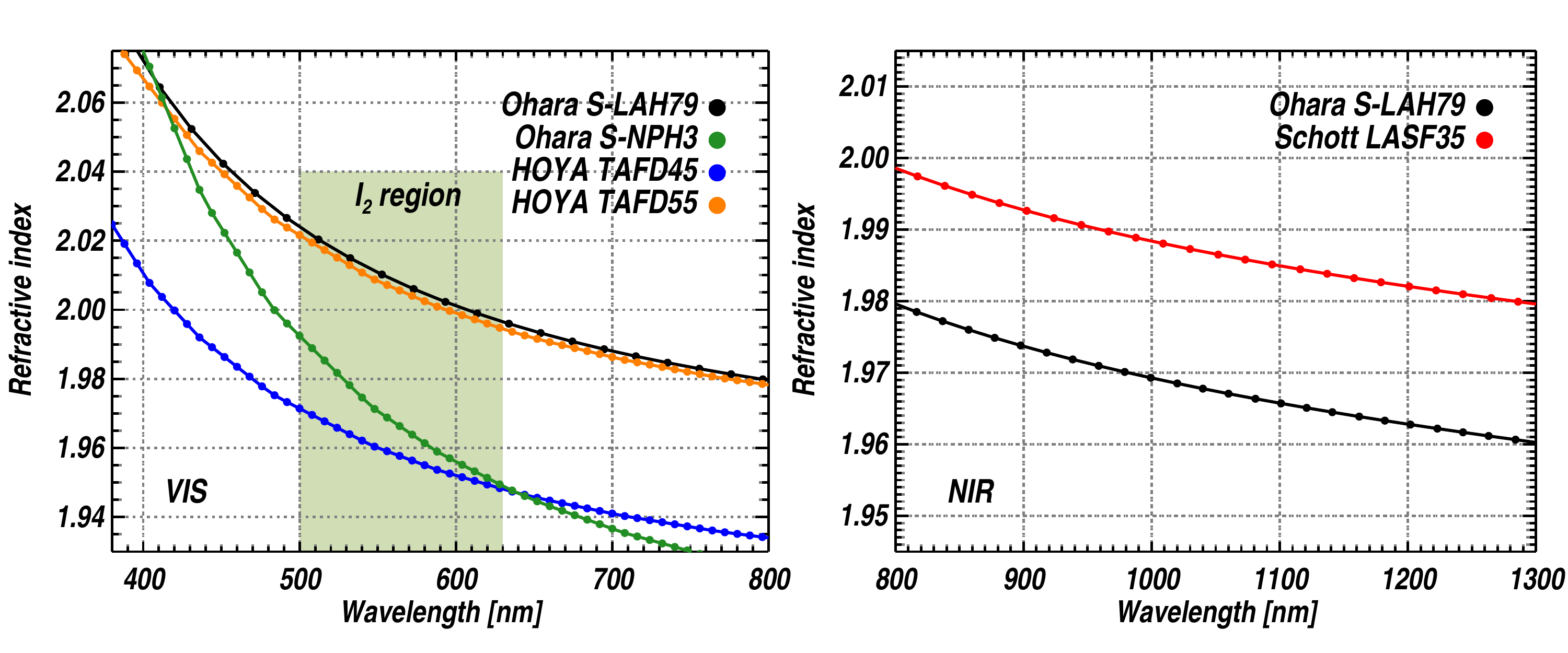}
\caption{Refractive index profiles for a variety of high-index glasses in the visible (left) and near-infrared (right). The lens used in our double-scrambler device is made of Ohara S-LAH79 glass. The ideal glass for a ball scrambler has refractive index of $n\sim2$ across a desired wavelength range with low $dn/d\lambda$.}
\label{fig:ball_index}
\end{center}
\end{figure}

We considered two high-index glasses, S-LAH79 (Ohara, n$_d$=2.00330) and LASF35 (Schott, n$_d$=2.02204) but chose to test S-LAH79 for the prototype ball scrambler due to its availability from vendors in small quantities\footnote{This proved prescient since LASF35 has since gone out of production.}. The optimal choice between materials depends on the specific wavelength range of interest but both glasses are well suited for the HPF bandpass. However, there are a variety of recently available high-index glasses from several manufacturers that cover large wavelength ranges in the visible and NIR with relatively low dispersion (see Figure~\ref{fig:ball_index}). When selecting the appropriate high-index glass for a particular application, both the refractive index and internal glass transmission must be considered to maximize efficiency. The two primary glasses discussed here have high throughput in the NIR ($>$99\% for a 2 mm lens), but have significantly diminished efficiency at wavelengths lower than 400 nm. However, select glasses shown in Figure~\ref{fig:ball_index} (left) do have relatively high transmission ($>$97\%) in the 380 - 400 nm regime for lens diameters under 1 mm, and would be more appropriate for optical instruments.

The ideal size of the ball ($D$) is calculated using Equation~\ref{eq:size}, as a function of the refractive index of the glass ($n$), the diameter of the input beam or fiber ($d$), and the input numerical aperture (NA). Our system is designed for a 300 $\upmu$m fiber fed at $f/3.65$ at the telescope focal plane, implying an ideal ball lens diameter of 2.15 -- 2.17 mm for our wavelengths. For concept verification purposes, we used more readily available 2 mm S-LAH79 ball lenses.  For optimum coupling and efficiency, the expected focal ratio exiting the fiber incident on the lens, rather than the native telescope focal ratio, should be used to select the size of the ball. This ensures that focal ratio degradation (FRD) and losses due to the lens are minimal.

Alignment of the ball between contiguous fiber faces is critical to producing a high efficiency scrambler. Classical double scramblers suffer throughput losses of 20-30\%, and an important aspect of our design is to substantially decrease this light loss. While concentricity between the lens and fiber faces is key, it is also necessary to control the distance between ball and fiber tip. With BFL ranging between 10 -- 20 $\upmu$m for our ball diameter and material, we essentially want the fiber butt-coupled to the ball. We developed two mounting methods to minimize coupling losses between the fibers and simplify the alignment process. Both devices yield 85-87\% efficiency, as measured using NIR fiber lasers, with minimal alignment effort using anti-reflective (AR) coated ball lenses (see \S\ref{sec:efficiency}).

\subsection{FC connector prototype}
\begin{figure}
\begin{center}
\includegraphics[width=3.4in]{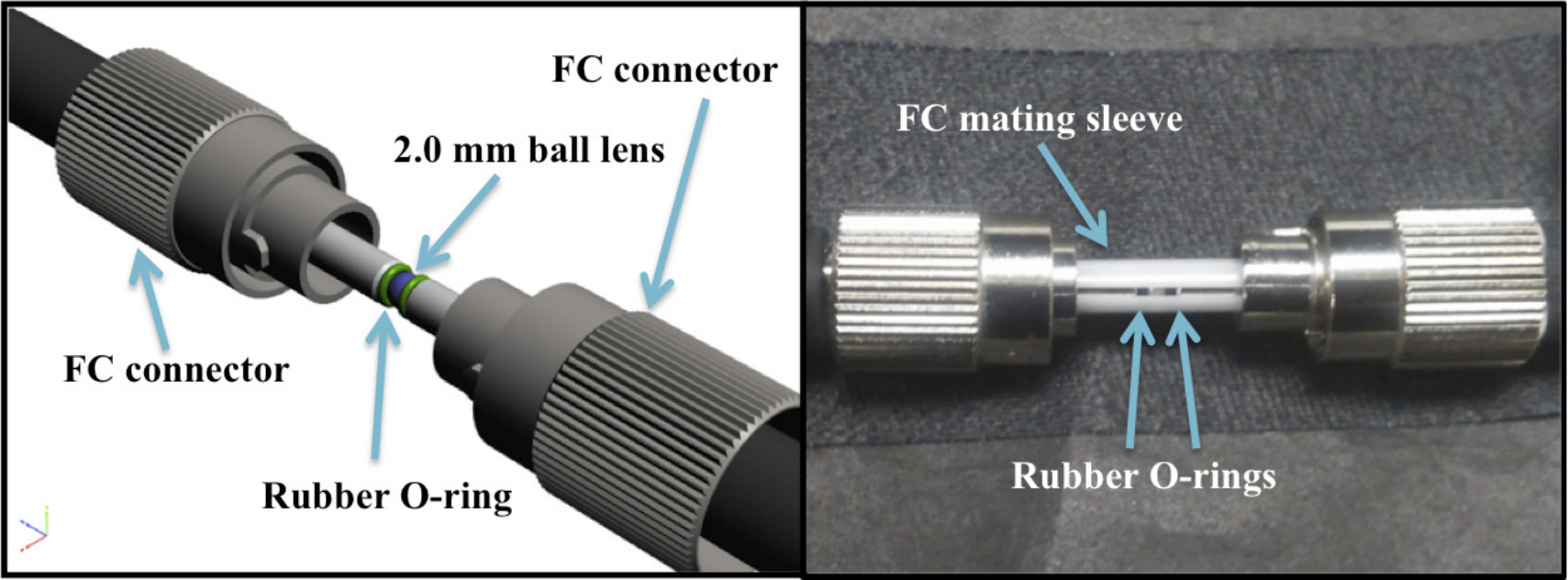}
\caption{Left: Solidworks model of FC connector ball lens double scrambler. The junction is held together by a commercial ceramic mating sleeve (not shown). Right: An assembled ball lens double scrambler with octagonal fiber cables on either side.}
\label{fig:align}
\end{center}
\end{figure}

In anticipation of a custom-built alignment piece, we implemented a simple and inexpensive design whereby the ball lens is held between two small rubber O-rings (Figure~\ref{fig:align}). The outer diameter of the O-rings (2.5mm) matches the ferrule size of the commercially standardized FC/PC fiber connectors used on our test fibers, and the entire junction can be placed within an off-the-shelf high-tolerance ceramic split sleeve of inner diameter 2.5mm. The lens is naturally kept along the central axis of the fibers by the pressure applied by the mating sleeve. In spite of the simplicity of this design, we are able to demonstrate efficiencies close to the theoretical maximum for a perfectly aligned system (\S\ref{sec:efficiency}) with this coupling scheme. This device was used for the the scrambling measurements presented in \S\ref{sec:results}.

\subsection{V-groove prototype}
\begin{figure}
\begin{center}
\includegraphics[width=3.4in]{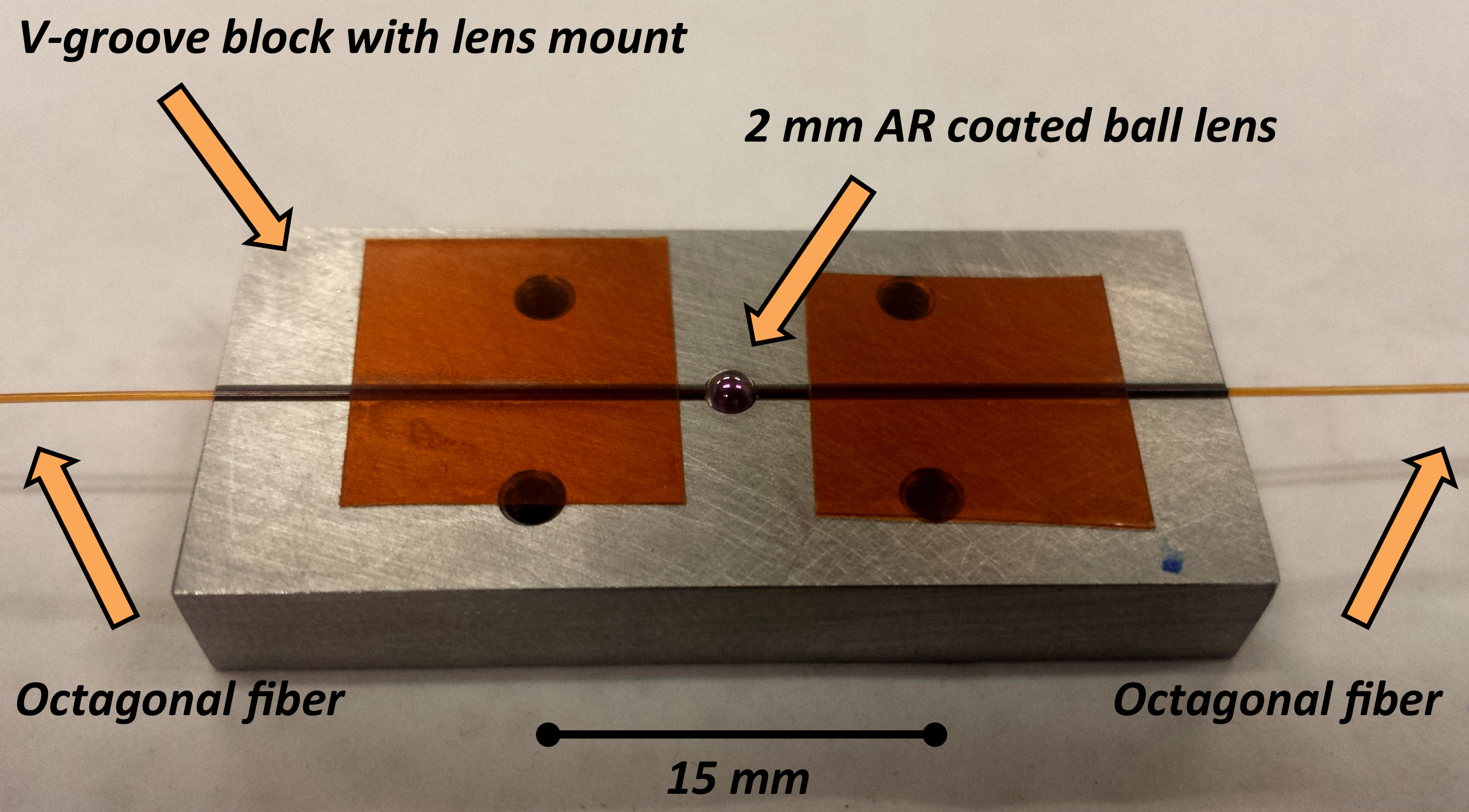}
\caption{V-groove mount for ball lens double scrambler. The ball lens is placed in a custom sized bevel in the center of the v-groove block. Polished octagonal fibers are guided inside the v-groove and held in direct contact with the lens on both side.}
\label{fig:vgroove}
\end{center}
\end{figure}

Our second prototype device consists of a custom manufactured, monolithic v-groove block to mount both fibers and the ball lens. The fibers are placed in a high precision v-groove, while the ball is placed in a drilled mount at the center of the stainless-steel block. The block is ground to size such that three faces can be used as reference indexing surfaces to within $\sim$12 $\upmu$m. Much attention was payed to maintaining the central axis of each element (input fiber, lens, output fiber) to high accuracy ($<$3 $\upmu$m). This level of accuracy was achieved by taking repeated surface measurements of the the machined block at several stages of the machining process using a high precision SmartScope measurement microscope. A tapered mounting pocket was machined on the block surface to secure the lens on the center of the block face. The SmartScope was then used to measure the central axis of the mounted lens relative to the polished block surface. The microscope measurements dictate the location and depth of the fiber v-groove relative to the block face, ensuring concentricity between the mounted lens and the fibers. The v-groove was etched using a high precision electric discharge machine. Post machining, the central axis of the lens and fibers were measured to be concentric at the measurement precision of the SmartScope ($\sim$3 $\upmu$m). An image of the completed v-groove block is shown in Figure~\ref{fig:vgroove}. Polished octagonal fibers are placed in the groove on either side of the lens and attached to the block with Kapton tape. 

\section{Scrambling Measurements}

For laboratory experiments we measure centroid values as a a proxy for scrambling performance. While this is a reasonable metric for near-field scrambling gains, since near-field illumination centroid drifts will directly manifest as velocity drifts in the focal plane, the direct effect of fiber far-field variation on velocity precision is not as easily quantifiable. We quantify the degree of far-field output variations for both near and far field input illumination changes, but do not formally calculate the effect of far-field variations on instrument velocity precision since this calculation is unique to a given instrument. This can be done with accurate ray-tracing of the fiber far-field through the spectrograph optics for a given instrument to set scrambling gain requirements \citep{Sturmer:2014}.

\begin{figure*}
\begin{center}
\includegraphics[width=3.4in]{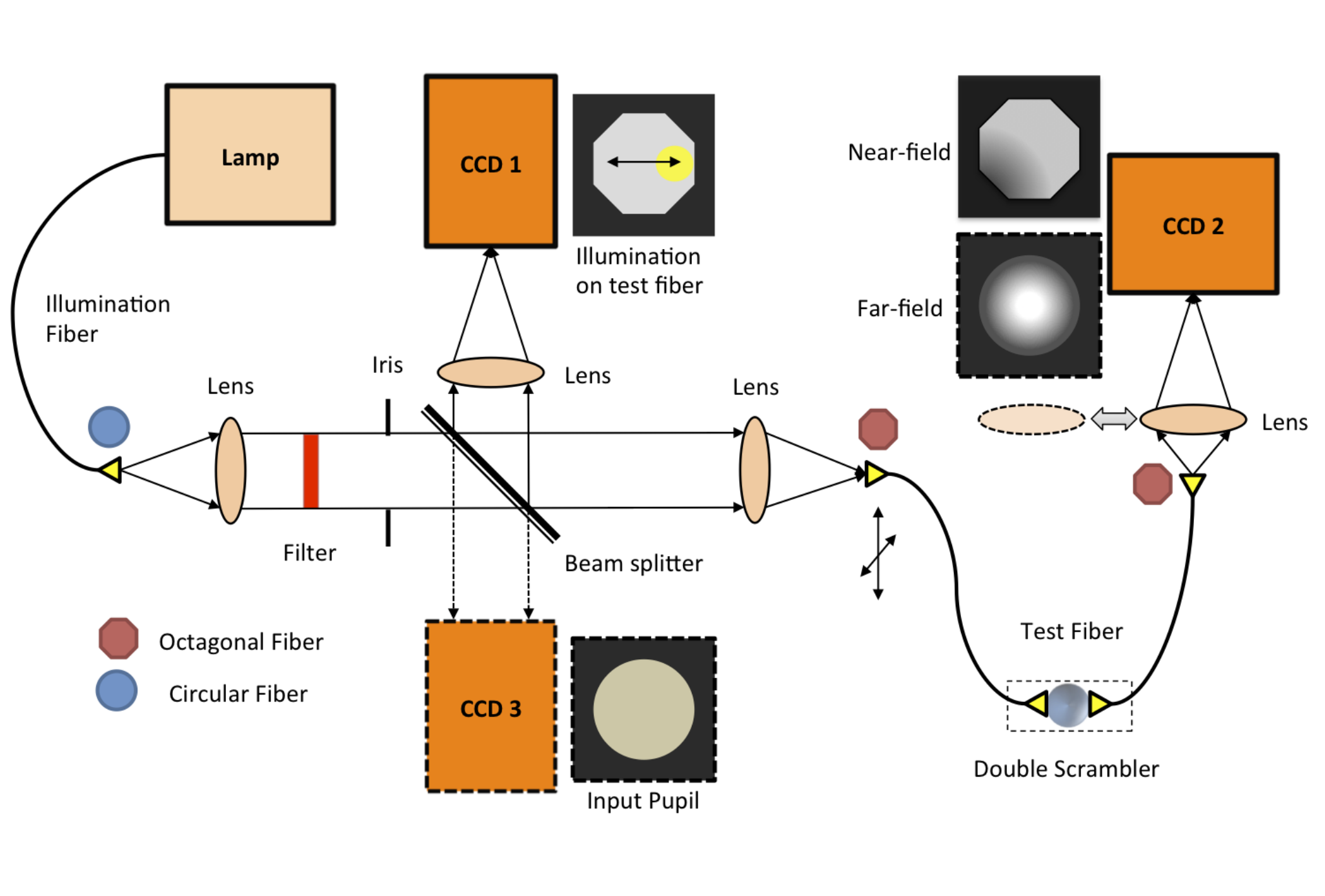}\includegraphics[width=3.4in]{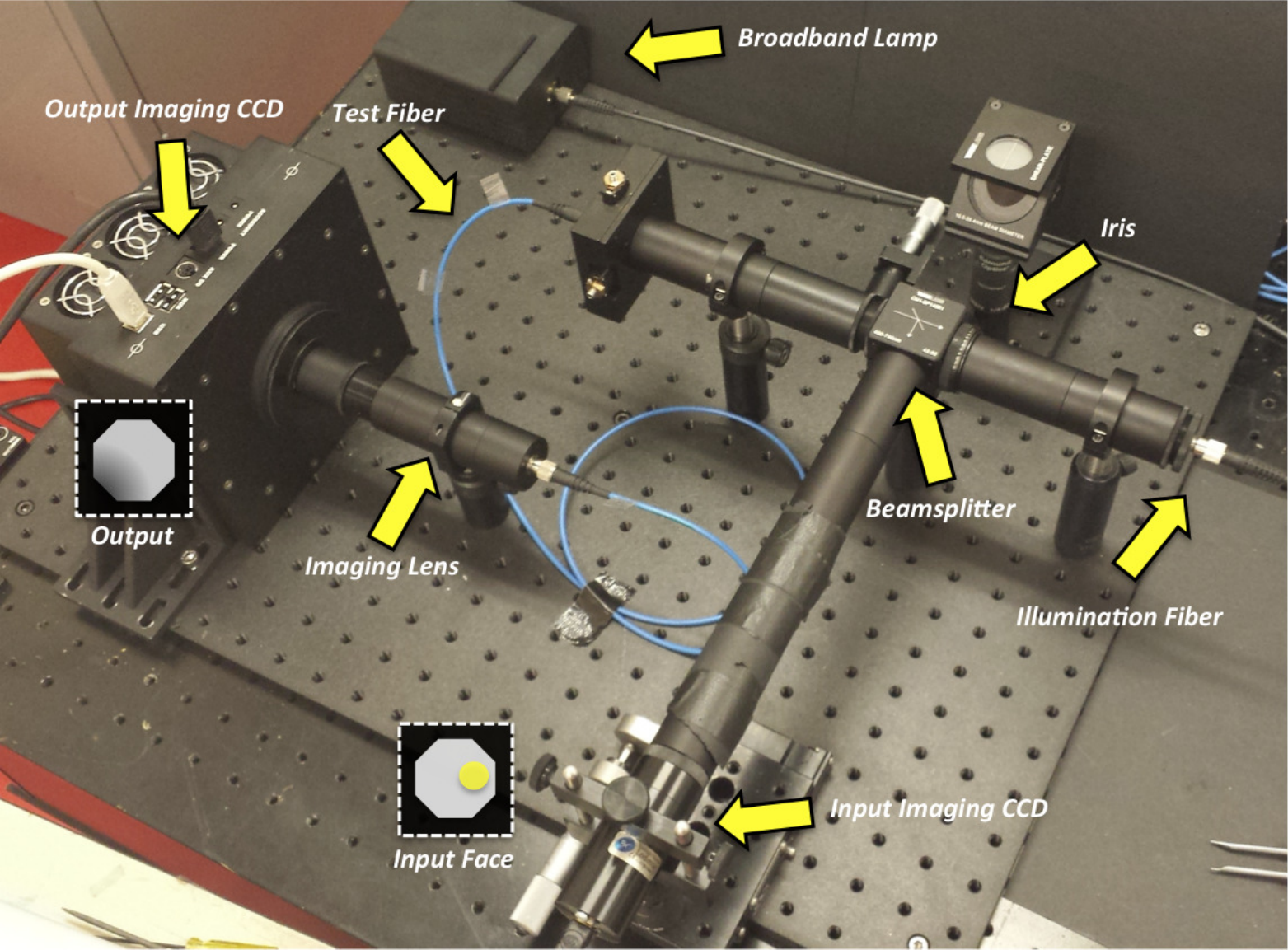}

\caption{Left: Schematic of fiber scrambling measurement test-stand. Light from a broad-band lamp is coupled into a 50 $\upmu$m illumination fiber. The output of the 50 $\upmu$m fiber is imaged onto a 300 $\upmu$m test fiber on an XY translation stage through a NIR filter. An iris is used to precisely control the input numerical aperture of the test fiber. The output of the test fiber is then imaged on a laboratory CCD. A lens is added to switch between near and far field imaging modes. A separate camera system is used to image the input face of the test fiber and track the location of the 300$\upmu$m fiber relative to the fixed illumination fiber. A third detector can be inserted to measure the input pupil incident on the fiber. Right: Image of scrambling gain measurement apparatus. The majority of components are connected through a series of lens tubes to reduce differential mechanical drift.}
\label{fig:test_stand}
\end{center}
\end{figure*}

\begin{figure}
\begin{center}
\includegraphics[width=3.3in]{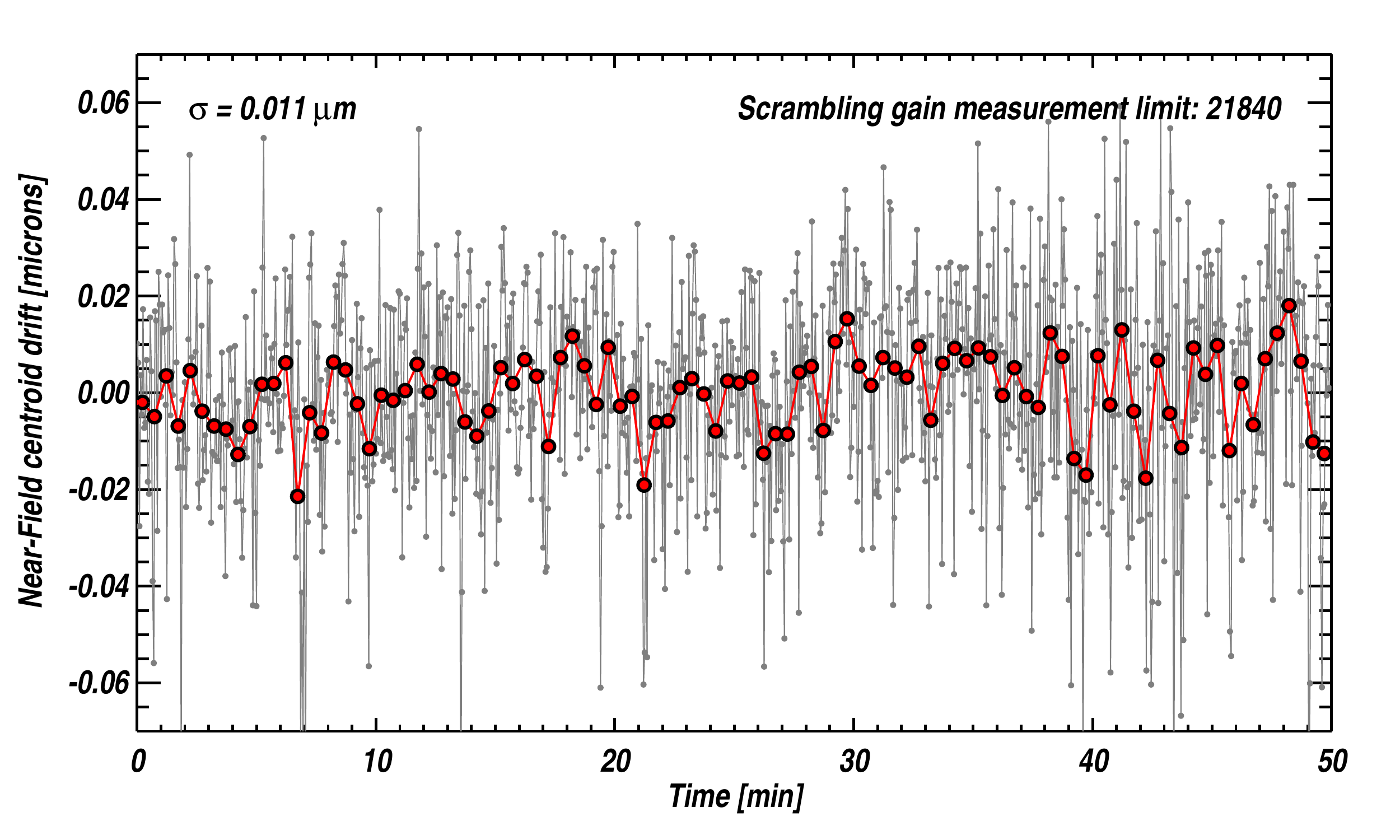}
\caption{Stability of test fiber output on measurement test stand. The input illumination was not moved during measurements. Centroid offset values for individual near-field images are shown in gray. The vertical axis is the measured centroid drift in units of microns on the fiber face. 30 second binned centroid values are shown in red.  The binned RMS stability of the fiber image is 0.011 $\upmu$m over tens of minutes of uninterrupted measurements, allowing for reliable measurements of scrambling gains $<$20,000.}
\label{fig:scrambling_stability}
\end{center}
\end{figure}

To be sensitive to scrambling gains larger than $\sim$10,000, a  compact laboratory test-stand was designed to maximize measurement stability. An overview of the experimental setup used to measure scrambling gain for a variety of fiber configurations is shown in Figure~\ref{fig:test_stand}. Light from a 50 $\upmu$m illumination fiber is focused onto a 300 $\upmu$m test-fiber at a speed comparable to the HET telescope delivery optics ($f/3.65$). A broadband Quartz lamp is used for illumination to minimize modal noise effects. We found speckle-noise to be the dominant source of measurement error when using narrow-band laser sources. A NIR filter is used to restrict the wavelength coverage to the HPF bandpass (0.8 -- 1.3 $\upmu$m). All optical elements and stages are connected through a series of lens tubes that are all bolted onto a common 2' $\times$ 2' optical breadboard. This ensures differential mechanical drift between optical components is reduced. The output of the test fiber is imaged over roughly 500 pixels of a 1024 $\times$ 1024 pixel laboratory CCD.  A second 400 $\times$ 400 pixel CCD is used to image the fiber input face and measure the location of the illumination spot relative to the fiber. A third detector can be placed in the pupil plane to measure the input pupil of the test fiber (see Figure~\ref{fig:test_stand} left). Our test apparatus does not allow for simultaneous measurement of both the near and far fields of the fiber output, though both have comparable measurement precisions. Over typical measurement time-scales (tens of minutes), our measurement apparatus is sufficiently stable to measure scrambling gains of $<$20,000 (see Figure~\ref{fig:scrambling_stability}). 

Optical fibers used for scrambling gain measurements are commercial 300 $\upmu$m Polymicro products. Octagonal fibers (FBP300/345/380) were connectorized and polished by C Technologies and circular fibers (FIP300/330/370) were terminated and polished by Polymicro. All patch cables were connectorized with standard 2.5 mm FC/PC connectors. Each cable used is roughly 2 m in length. Due to the high degree of FRD in our test fibers, the majority of the incident $f/3.65$ beam exited the test fiber at a faster $f/3.3$ beam. This faster beam happens to be well matched to the 2.0 mm ball lens used in our prototype device, minimizing any additional FRD due to the ball lens.

\begin{figure}
\begin{center}
\includegraphics[width=3.4in]{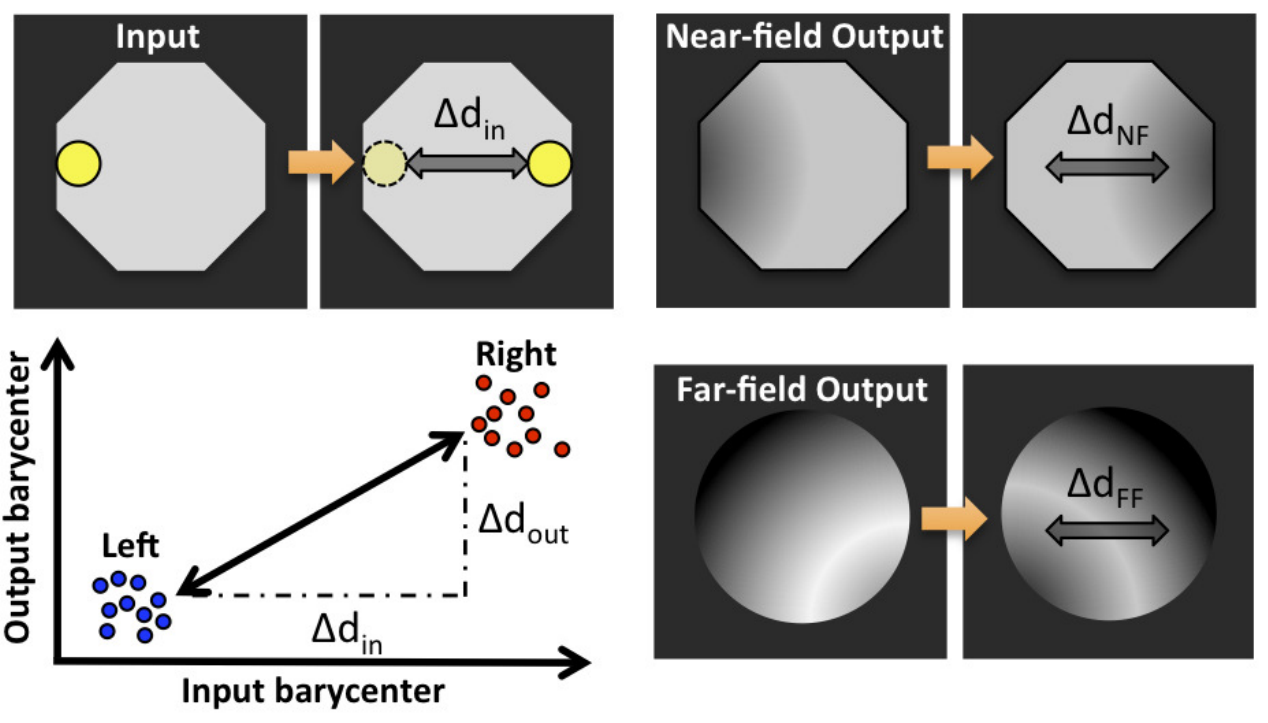}
\caption{Cartoon illustration of scrambling measurement method. The test fiber is translated across a fixed illumination spot and the fiber output is recorded at each position. The measured centroid shifts in the fiber near field (NF) and far field (FF) are compared to the distance translated at the input to calculate scrambling gain.}
\label{fig:scrambling_cartoon}
\end{center}
\end{figure}

Scrambling measurements were taken by shifting the test fiber input face across a fixed illumination spot and measuring the centroid offset in the fiber output in both near and far fields. The test fiber is shifted by $\sim$250 $\upmu$m across the fixed 50 $\upmu$m near-field illumination. Rather than taking measurements at intermediate points between two edges of the test fiber input, data are taken only at maximum displacement on the fiber face (see Figure~\ref{fig:scrambling_cartoon}). Multiple 3-s exposure images were recorded at each edge of the test fiber face and averaged. Dark frame and flat-field corrections were then applied to the averaged images. An image mask was applied to isolate flux coming from only the fiber face. 

The centroid of the output fiber illumination is then calculated in both spatial directions and compared to the location of the illumination spot to calculate a formal scrambling gain. ${\Delta}d_\mathrm{input}$ is the distance translated by the focused input illumination and ${\Delta}d_\mathrm{output}$ is the measured centroid shift in the fiber output (near or far field) along the axis that yields the largest measurable shift. Output centroid measurements at both illumination positions are used to calculate individual scrambling gain estimates for both the near-field and far-field outputs. The quoted scrambling gains are averages of 10 point-to-point scrambling gain measurements.

This method has the benefit of producing the largest measurable output centroid shift over the shortest time-scale, thereby minimizing sensitivity to short-term optomechanical instabilities in the measurement apparatus. Measurements were repeated along several translation axes on the fiber face to verify results. Much attention was paid to make sure that the cladding of the test fiber was not illuminated, though no visible cladding modes were present even when the illumination was off the fiber core entirely. 

Note that, as mentioned previously, this measurement is only gauging the effect of near-field illumination variation on the near and far-field of the fiber output (i.e. the input pupil incident on the test fiber is not varying). In order to measure the effect of a varying input pupil on the fiber output, a variable pupil mask was inserted at the fiber input pupil and rotated. This experiment is meant to probe the extreme case of possible input pupil variation for the HET, though we do not calculate a formal scrambling gain for this measurement.

\section{Results}
\label{sec:results}
\begin{figure*}
\begin{center}
\includegraphics[width=1.42in]{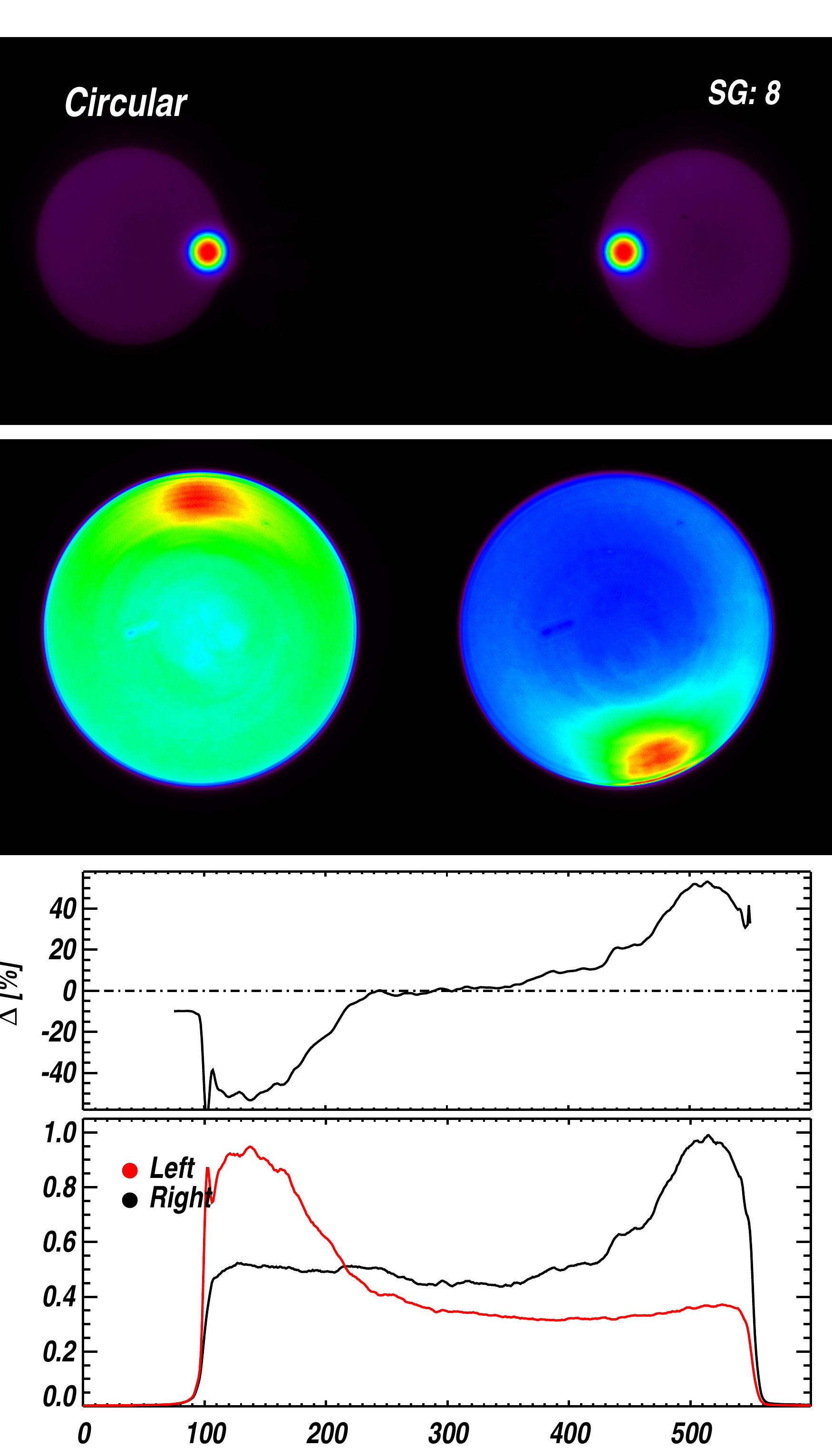}\includegraphics[width=1.42in]{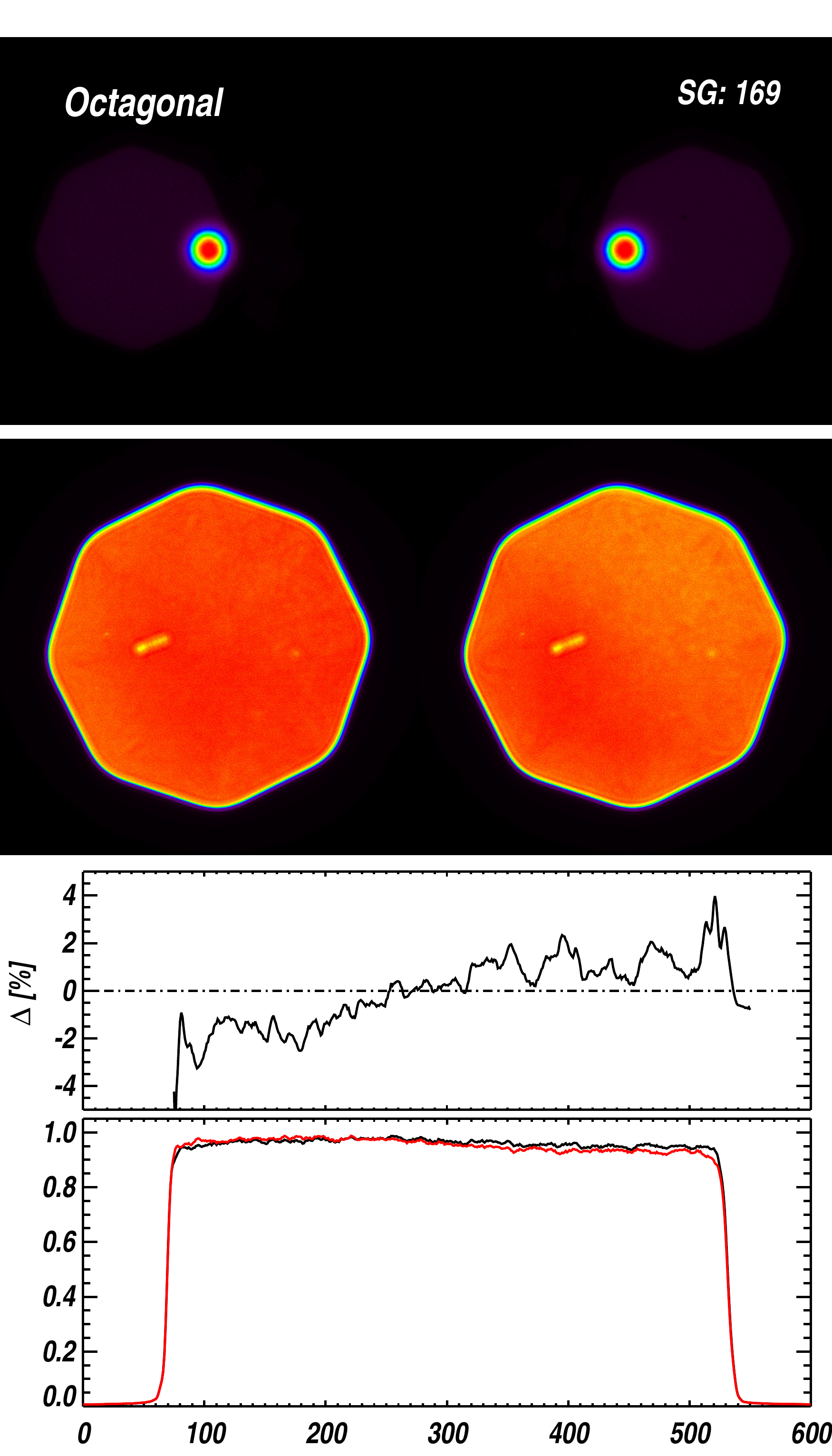}\includegraphics[width=1.42in]{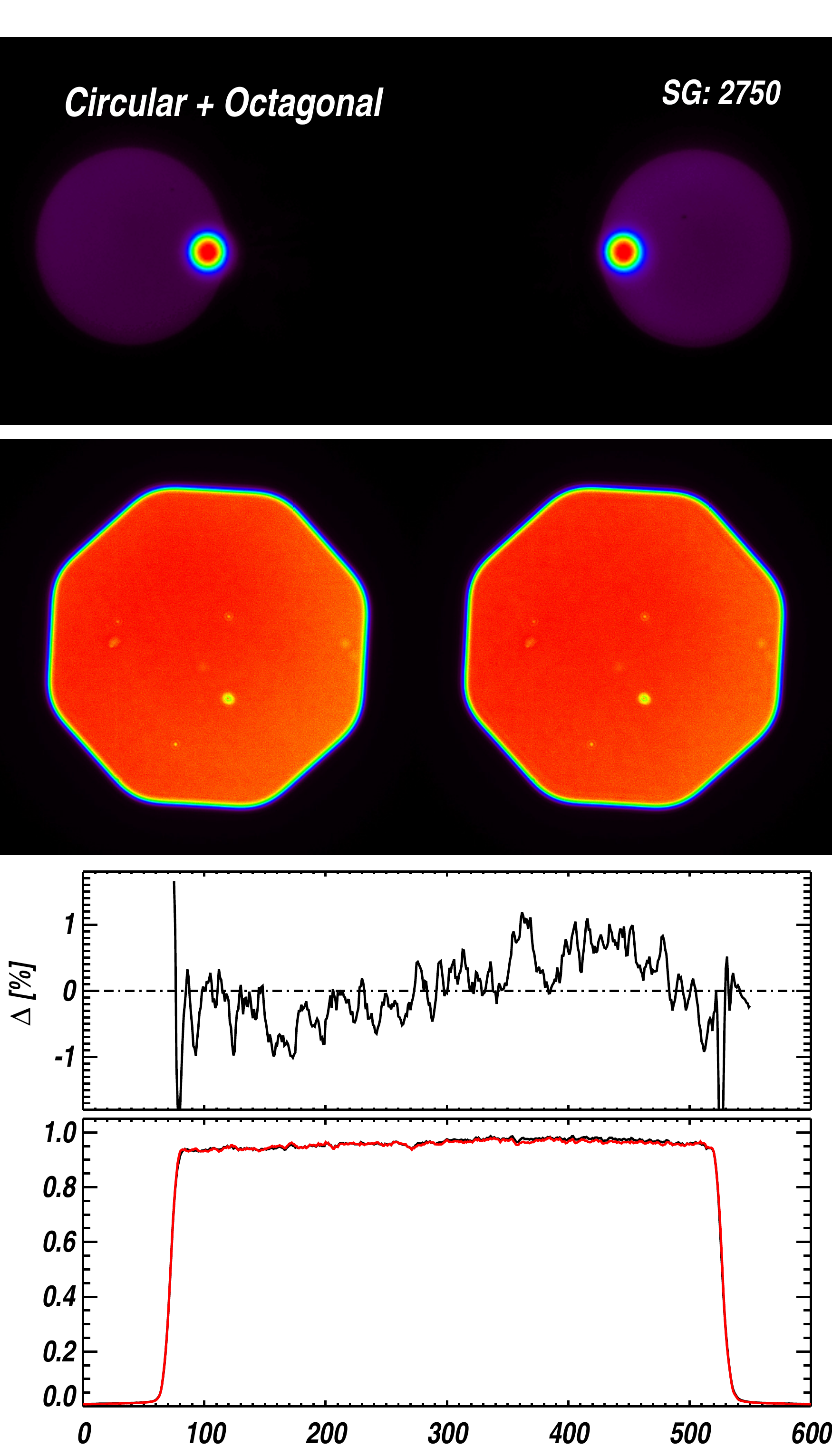}\includegraphics[width=1.42in]{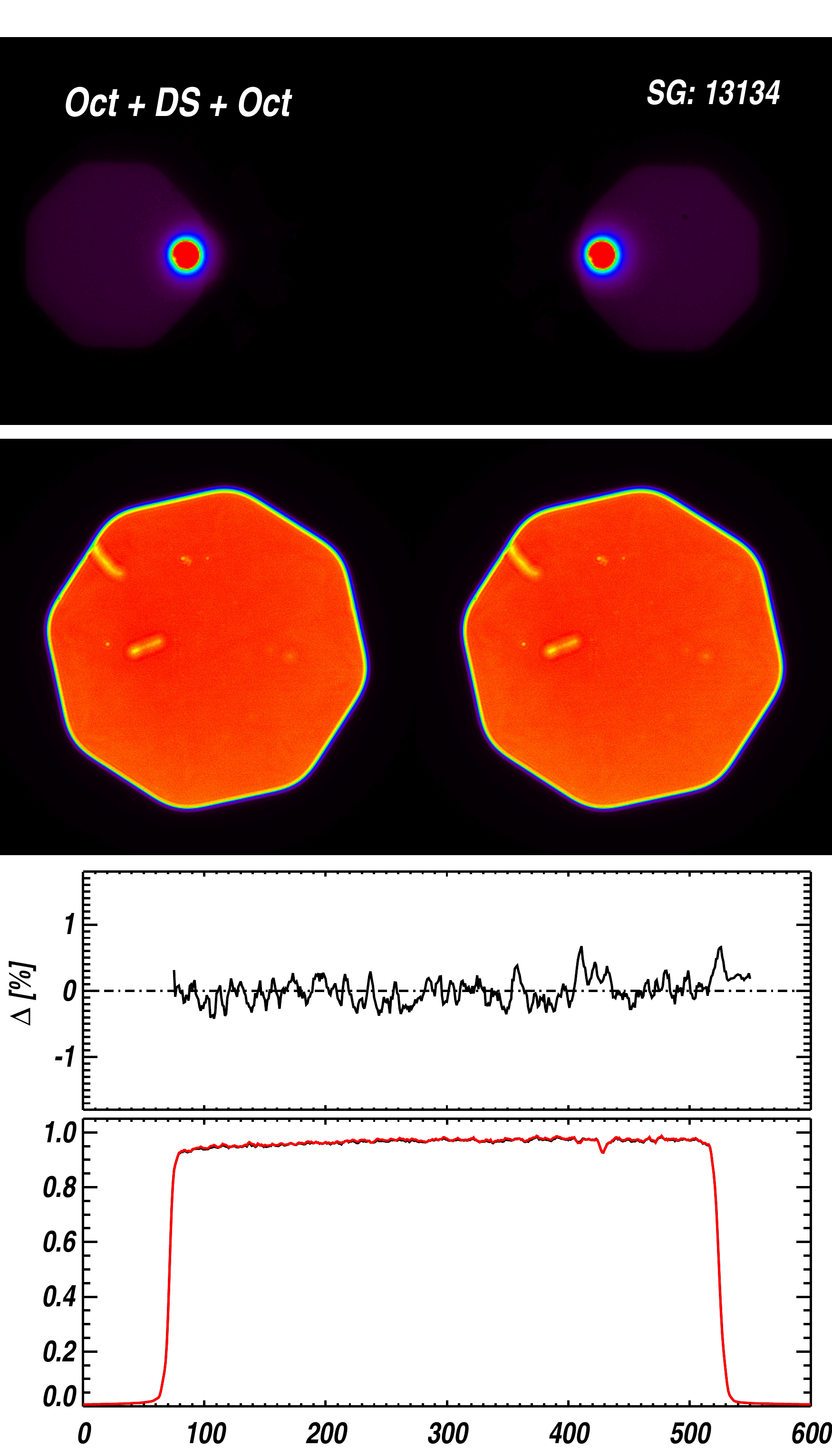}\includegraphics[width=1.42in]{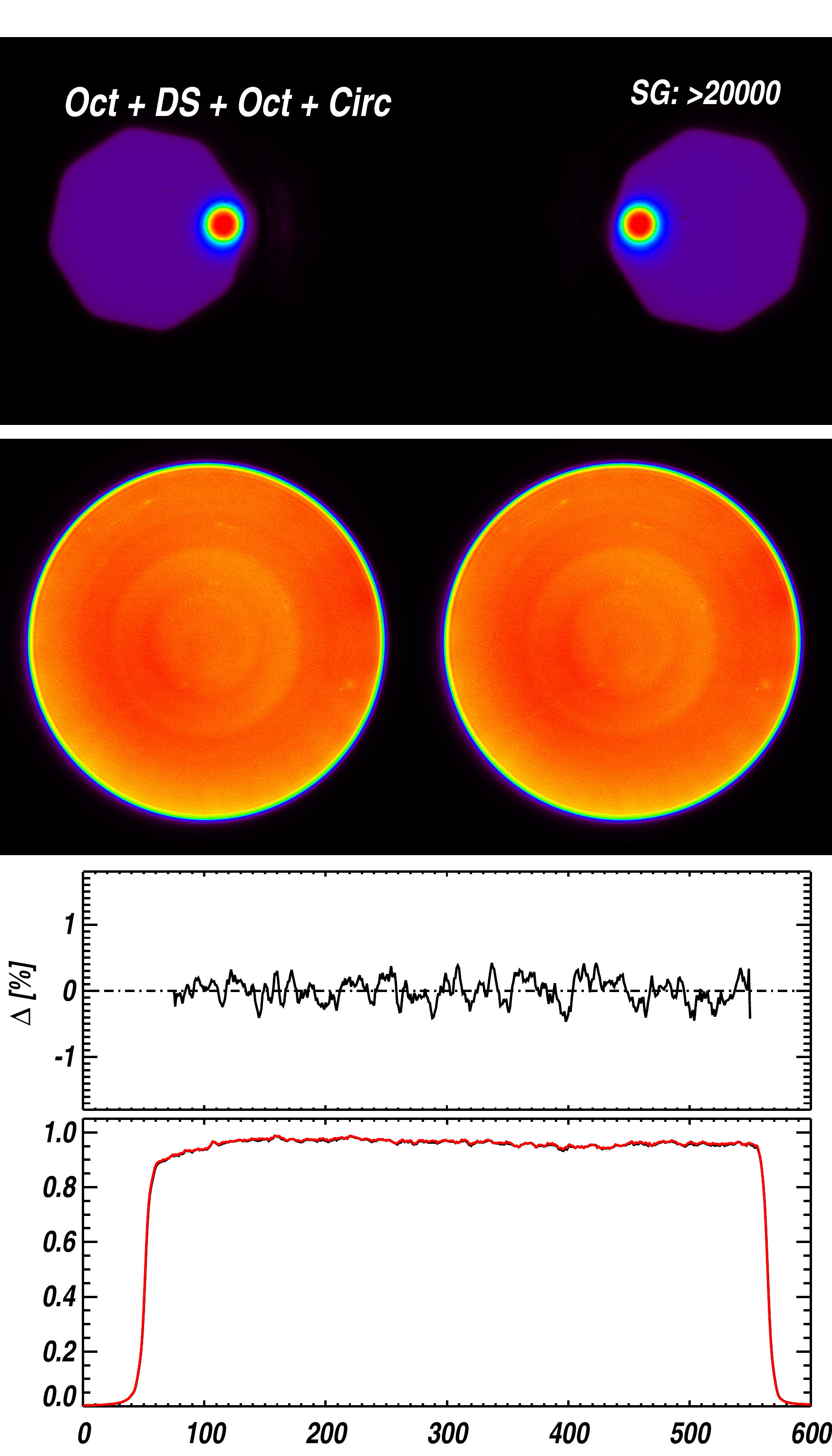}
\caption{Near-field scrambling measurements for different fiber configurations. Top panels show the location of the input face of the test fiber relative to illumination spot. The near-field output images are shown in the middle panels. Scrambling gains (SG) are calculated by dividing input illumination shift by output centroid drift. Bottom panels show the 1D intensity distribution along the center of the fiber for the two different illumination positions and the difference ($\Delta$, in percentage) between the two 1D profiles. Note the vertical axis scales differ between individual difference plots. The full double-scrambler system, which includes octagonal fibers (coupled by a ball lens) and a circular fiber, produces a stable and uniform output illumination that is significantly desensitized to input illumination variations.}
\label{fig:NF_ex}
\end{center}
\end{figure*}

\begin{figure*}
\begin{center}
\includegraphics[width=1.42in]{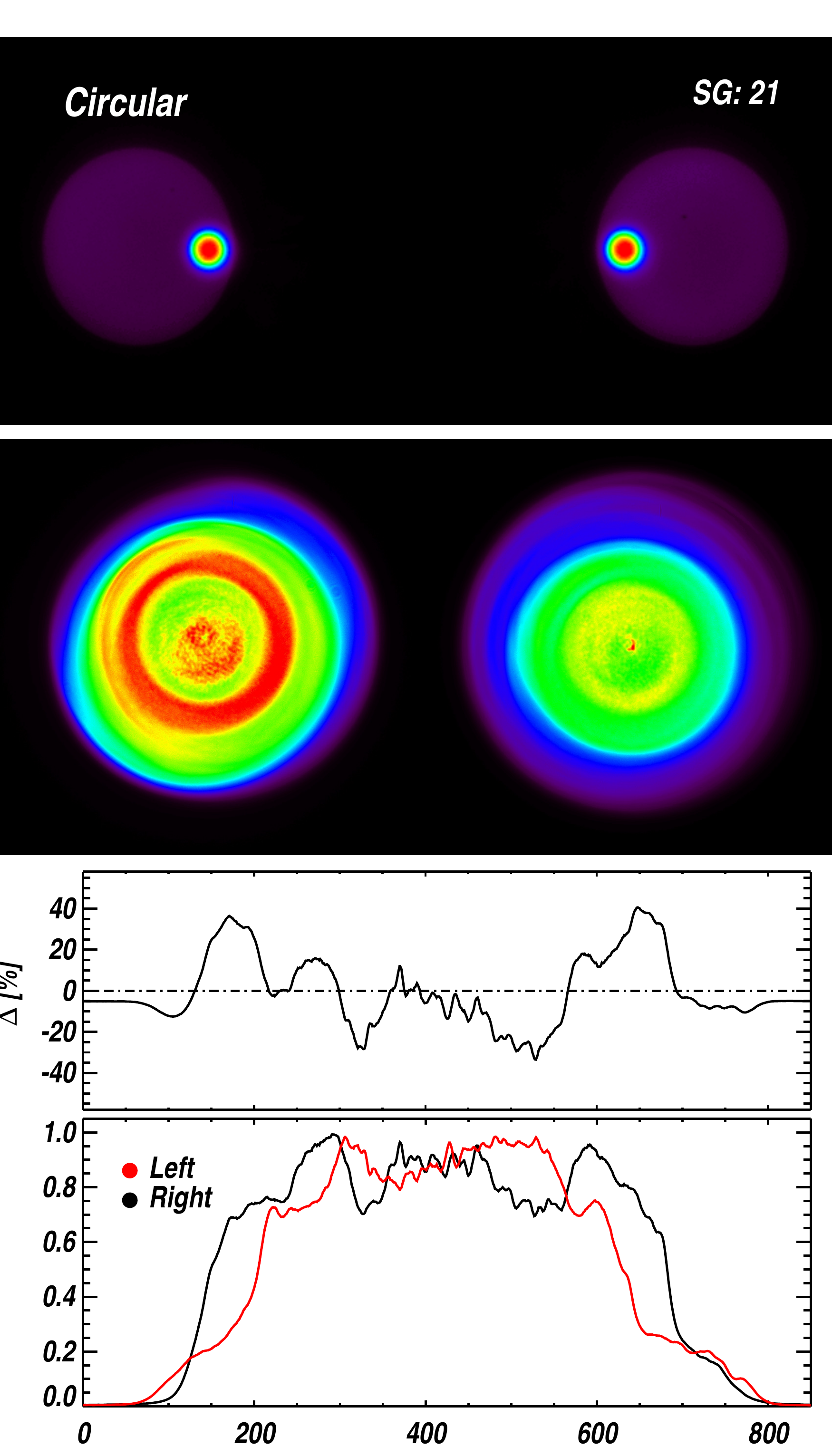}\includegraphics[width=1.42in]{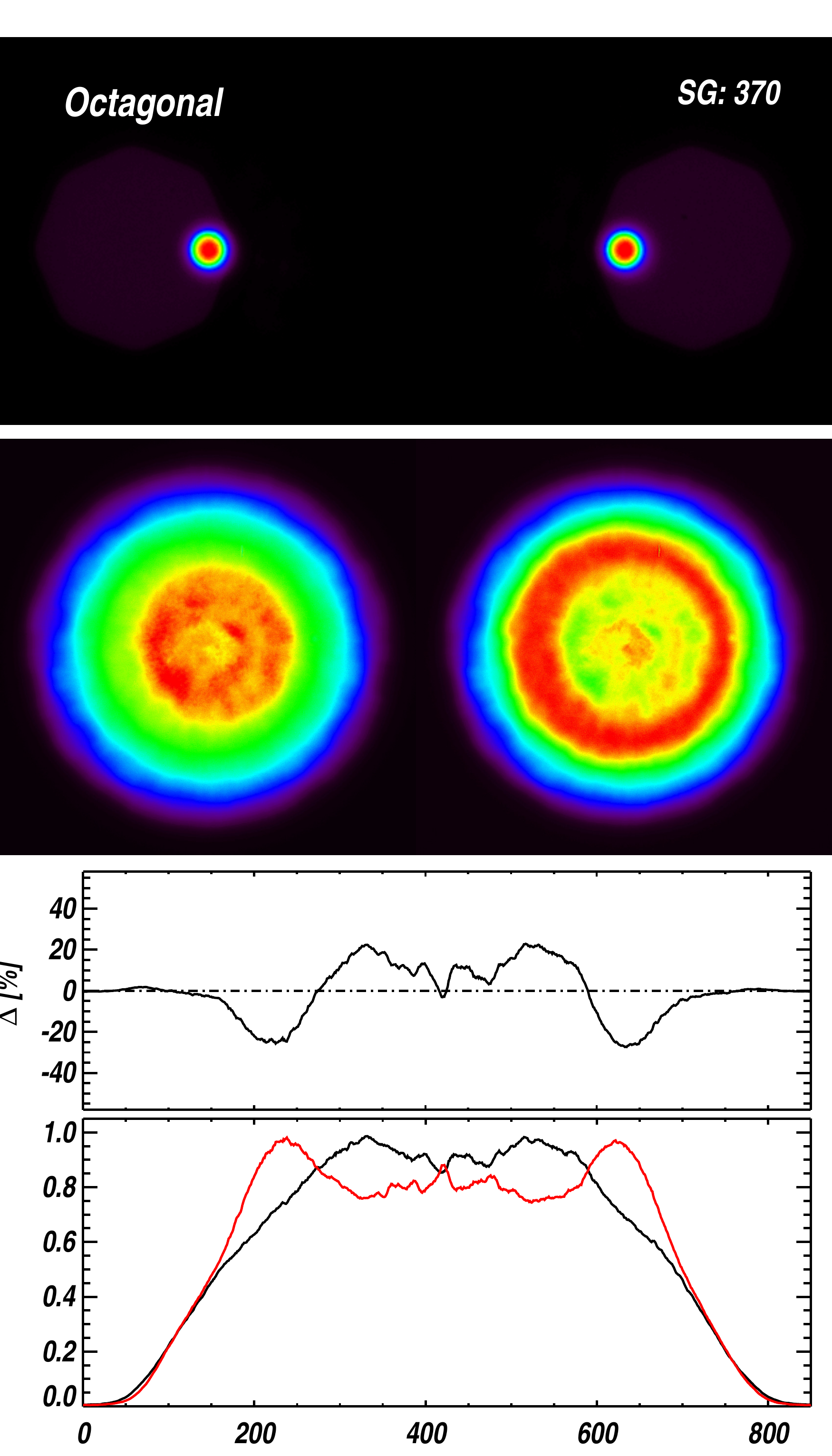}\includegraphics[width=1.42in]{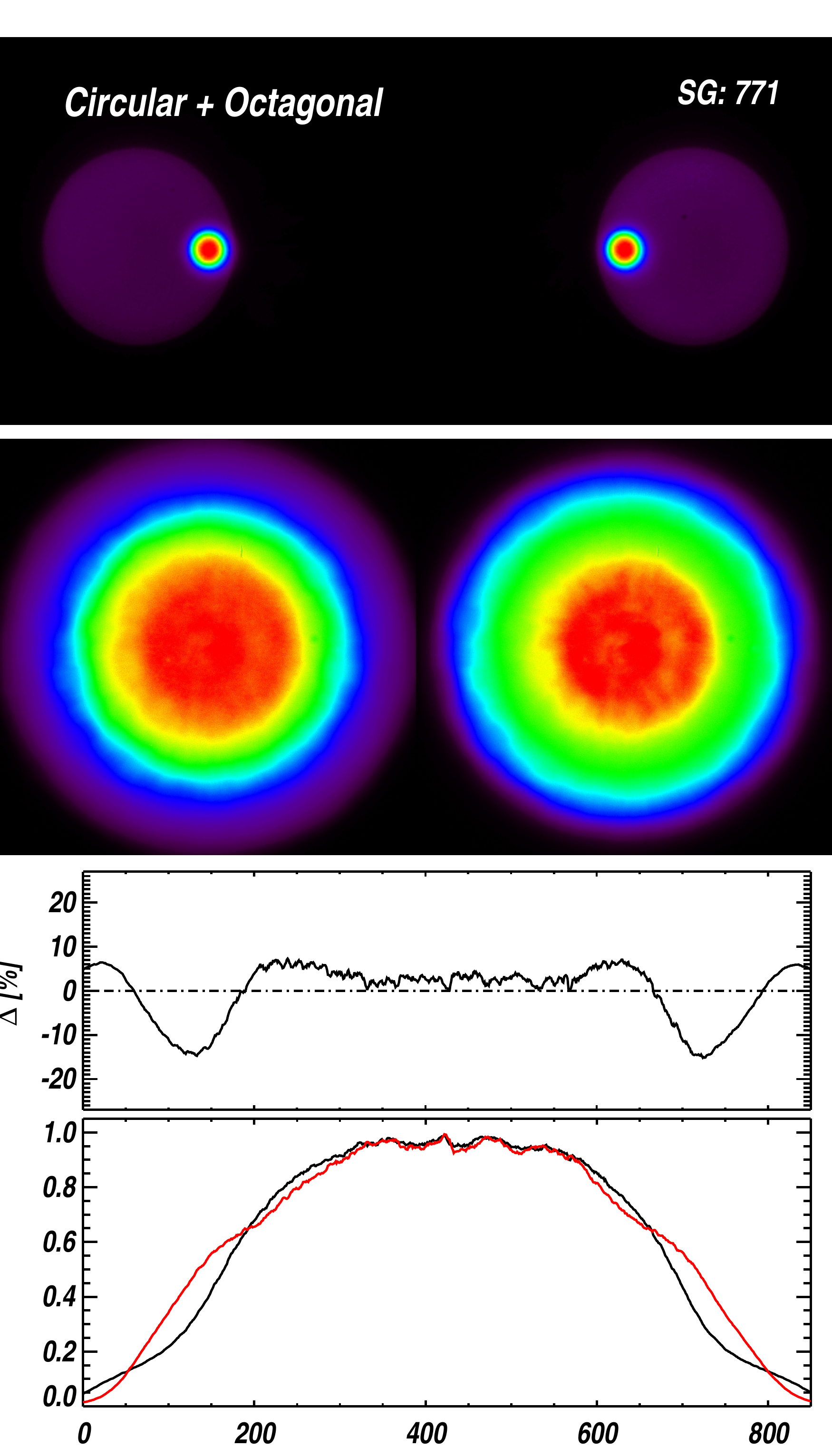}\includegraphics[width=1.42in]{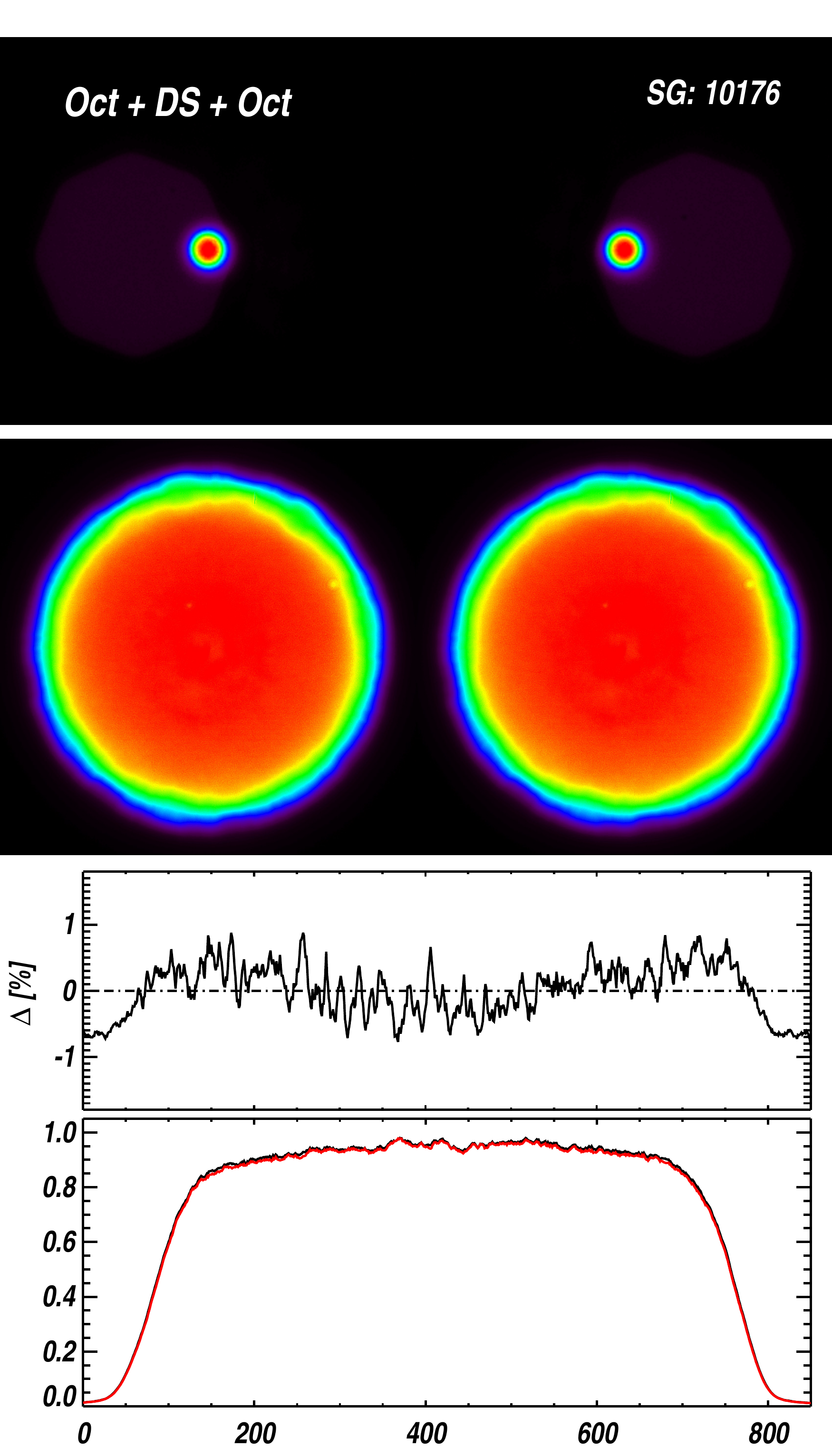}\includegraphics[width=1.42in]{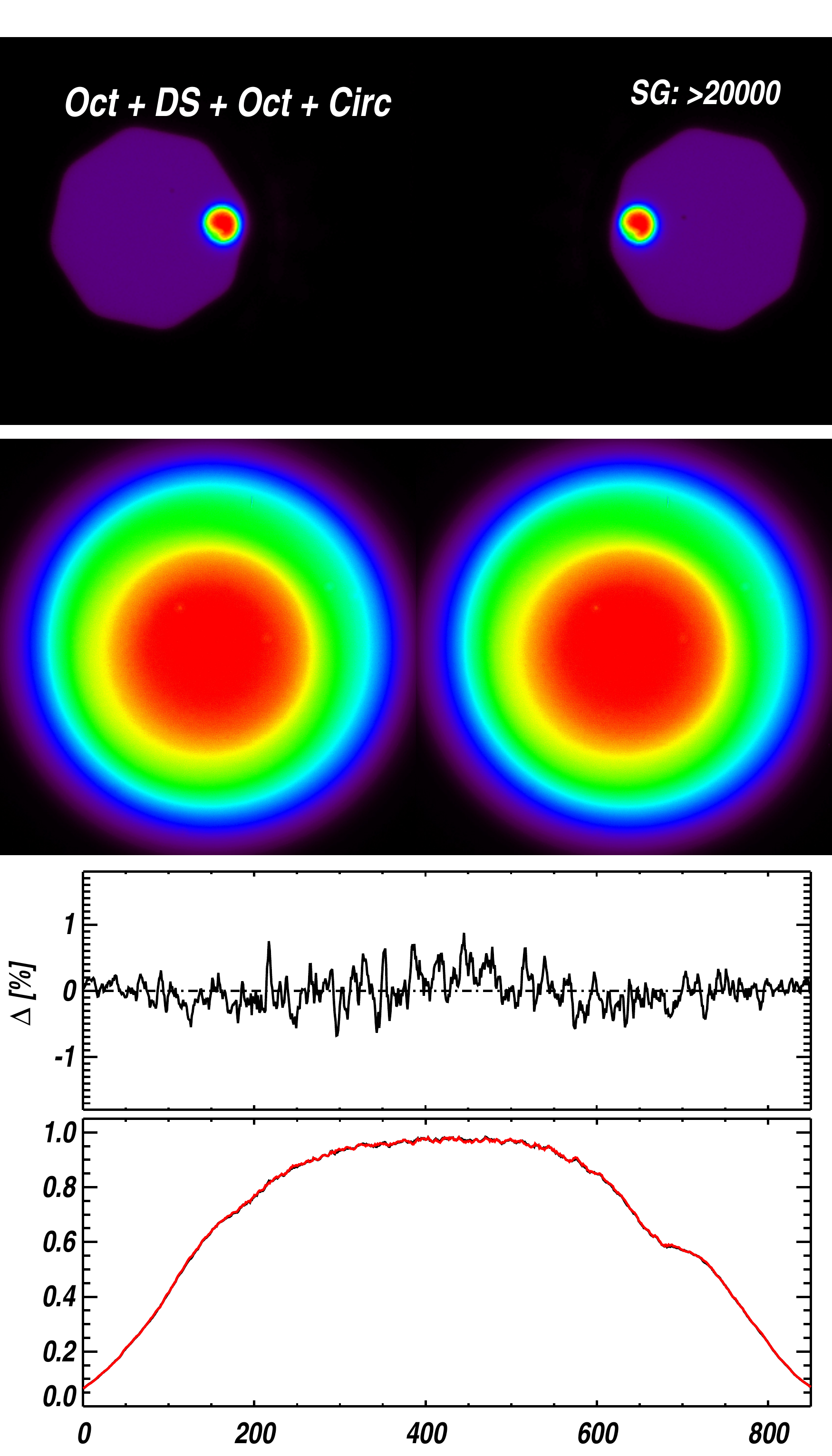}
\caption{Effect of near-field variations on the fiber far-field output for different fiber configurations. The addition of the double-scrambler significantly improves the far-field uniformity and stability.}
\label{fig:FF_ex}
\end{center}
\end{figure*}

Laboratory measurements of scrambling gains for a variety of fiber configurations are shown in Figures~\ref{fig:NF_ex} \& \ref{fig:FF_ex}. Measured scrambling gains are listed in Table~\ref{tab:results} for both near and far field outputs. Circular fibers, as expected, yield relatively low scrambling gains. A single octagonal fiber yields modest scrambling gains of $\sim$100-400, consistent with previous measurements. We find the highest scrambling gain ($>$20,000) is achieved when using a combination of octagonal fibers, a double-scrambler, and circular fibers. In this configuration, the circular fiber is coupled to the end of the second octagonal fiber with a standard FC connector mating sleeve, The complete fiber train now includes, in order of light travel: first octagonal fiber, ball lens scrambler, second octagonal fiber, circular fiber. In future iterations of this configuration, individual fibers will be coupled with high tolerance FC connectors, or spliced together, to maximize throughput of the complete fiber train. This dictates the fiber feed configuration for HPF, since the HET has a particularly variable illumination pattern incident on the instrument fiber. As this measurement is limited by our laboratory apparatus, the true scrambling gain is likely higher. Nevertheless, this near-field scrambling gain translates to a $\sim$10~cm~s$^{-1}$ velocity shift for HPF, significantly better than our performance goal of 30 c\ms guiding-induced velocity error.

\begin{figure}
\begin{center}
\includegraphics[width=1.65in]{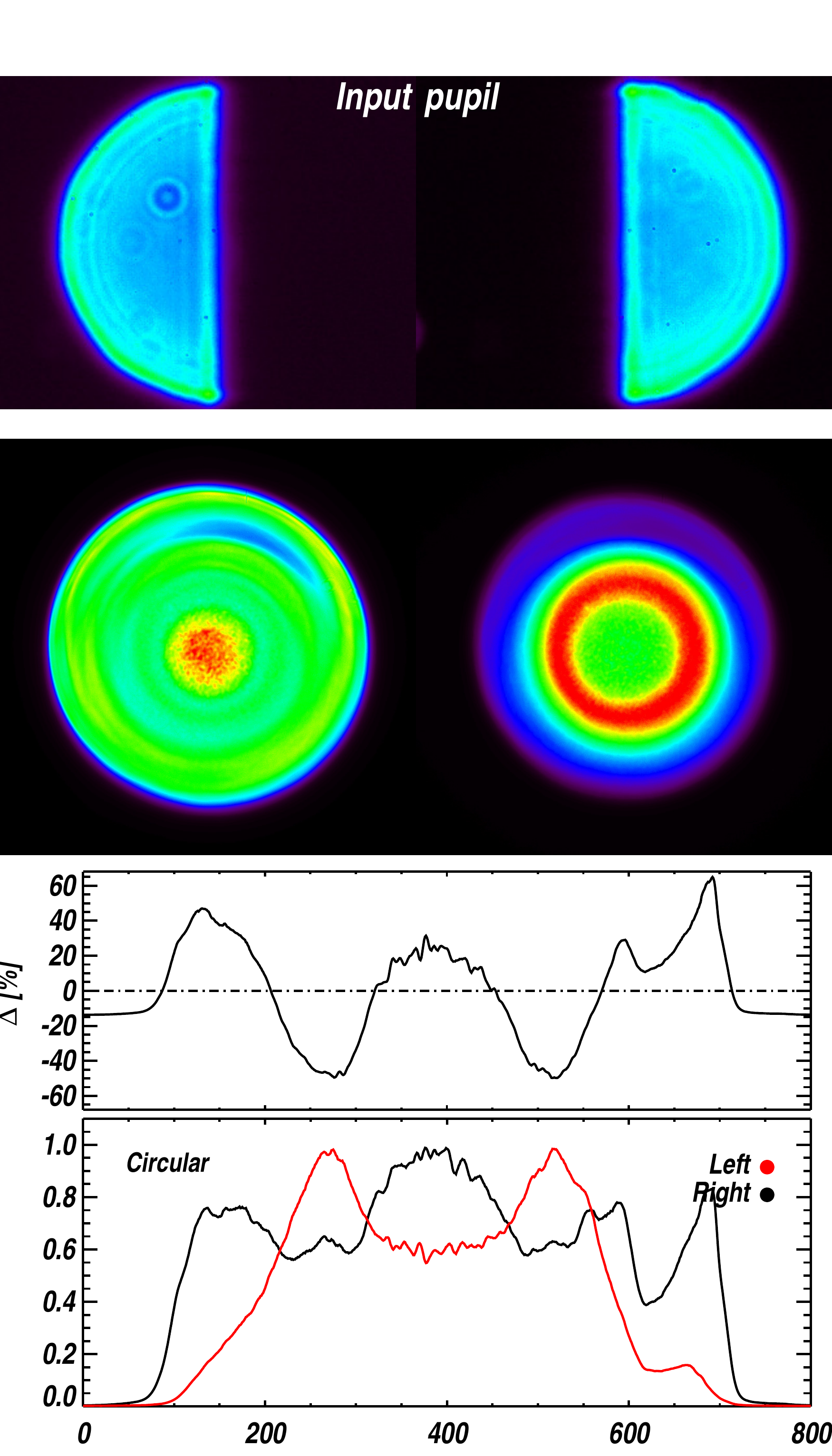}\includegraphics[width=1.65in]{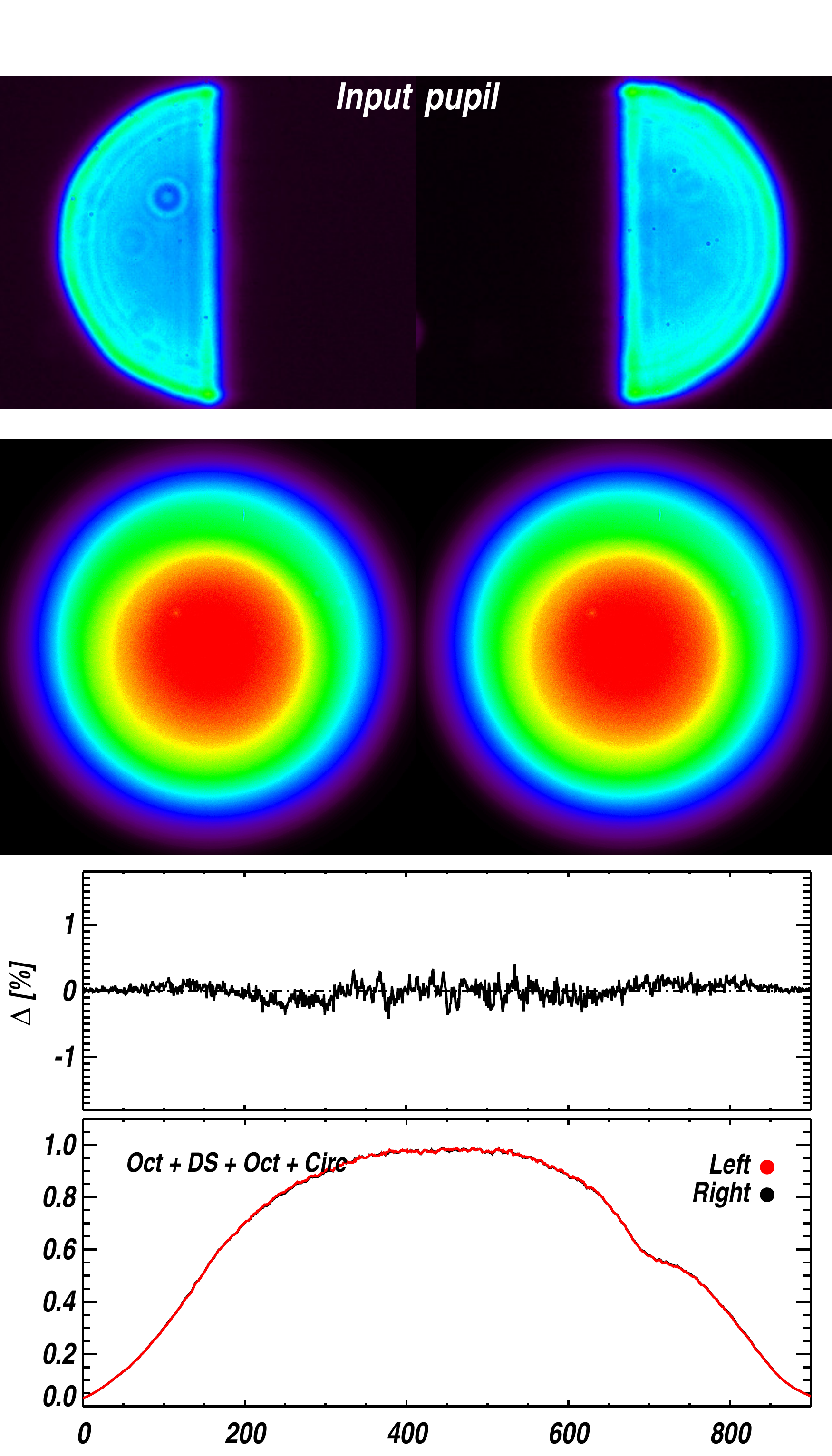}
\caption{Similar to Figures~\ref{fig:NF_ex} \& \ref{fig:FF_ex}. Effect of input pupil variations on fiber far-field output for a single circular fiber (left) and the octagonal-circular fiber double scrambler system (right). The top images show the input pupil to the test fiber. The middle panel shows the far-field output of the fiber system, and the lower plots show the 1D fiber cross-section profile. Even with extreme variation of the input pupil, the double-scrambler fiber system shows minimal change in output illumination.}
\label{fig:pupil_variation}
\end{center}
\end{figure}

We have not yet performed an extensive analysis to determine the velocity error associated with the observed output pupil variations of various fiber configurations, although we plan to do this in the future for HPF. We have already initiated experiments that mimic an extreme case of pupil variation. For these tests, the near-field illumination was kept static (i.e. the spot was fixed with respect to fiber face), but the far field was varied with a rotating aperture mask that obscures half of the input pupil. Figure~\ref{fig:pupil_variation} shows the measured far-field output of a single circular fiber compared to the fiber configuration that produces the best scrambling gains (octagonal fiber + double scrambler + octagonal fiber + circular fiber).  While the single circular fiber system changes dramatically to reflect the extreme pupil obscuration we impose at the input, our double scrambler fiber system shows minimal sensitivity to input pupil illumination variations. This is extremely promising for HPF, since the spectrograph will need to be dissociated as much as possible from the inconstant HET input pupil if the instrument is to meet the 1 \ms overall RV precision goal.

\begin{center}
\begin{deluxetable}{ccc}
\tablewidth{0pt}
\tablecaption{Measured scrambling gains for different fiber configurations.}
\tablehead{
\colhead{Fiber Configuration}  & \colhead{Near-field} & \colhead{Far-field}
}
\startdata
Circular Fiber	&	8	&	21 \\
Octagonal Fiber	&	169		&	370	\\
Circular + Octagonal		&	2750		&	771\\
Octagonal	 + DS + Octagonal	&	13,134	&	10,176 \\
Octagonal + DS + Octagonal + Circular	&	$>$20,000	&	 $>$20,000
\enddata
\label{tab:results}
\end{deluxetable}
\end{center}

\subsection{Overall Efficiency}
\label{sec:efficiency}

Apart from excellent scrambling gain, our compact double scrambler system yields high throughput when coupled with octagonal fibers. Classical double-scrambler systems yield measured efficiencies varying between 20-80\% (see Table~\ref{tab:eff}), limited mainly by the tight alignment tolerances and intrinsic reflective losses associated with uncoated optical systems. 

\begin{figure}
\begin{center}
\includegraphics[width=1.52in]{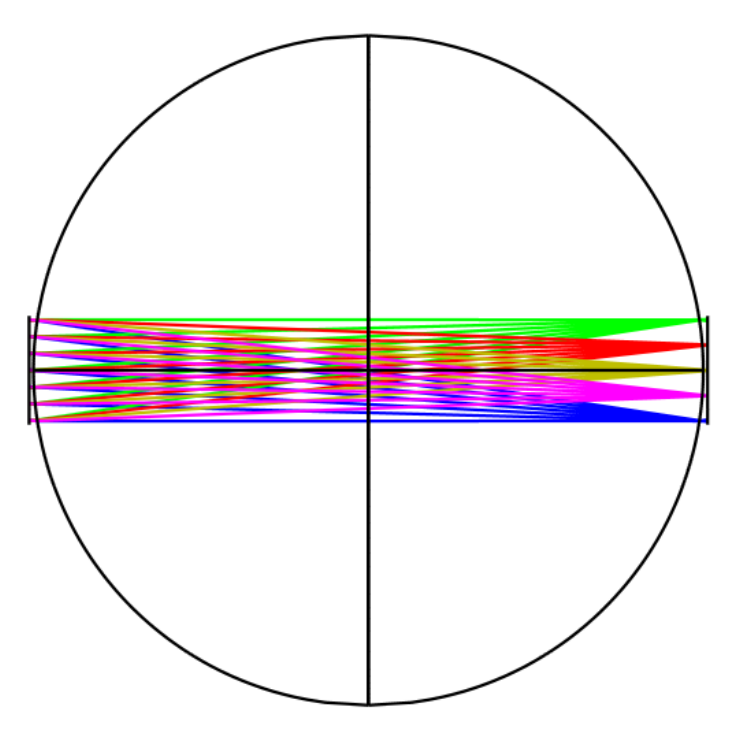}\includegraphics[width=1.9in]{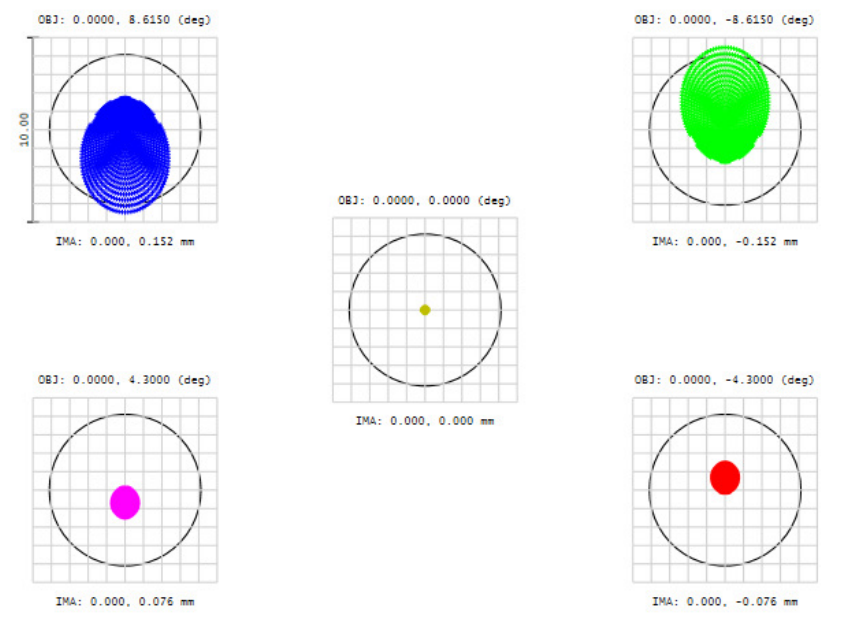}
\caption{Left: Ray trace of ball lens scrambler showing characteristic conversions from field-to-angle, where the pupil of the first fiber is projected onto the near field of the second fiber, and vice versa. Right: Spot diagrams for a variety of input field points are shown. Spots were generated at a wavelength of 1 $\upmu$m. The black circle is the expected diameter of the Airy disk (8 $\upmu$m).}
\label{fig:zemax}
\end{center}
\end{figure}

Consistent with Zemax simulations, we calculate our potential sources of throughput loss, namely Fresnel reflection, aberrations from the ball lens (see Figure~\ref{fig:zemax}), diffraction effects, and geometric losses arising from rotational misalignment between the two octagonal fiber faces on either side of the ball. A single ball lens element intrinsically has low spherical aberration compared to standard lenses of the same material \citep{Riedl:2001}. While the distance between the ball lens and the fiber tip can vary by $\sim$20 $\upmu$m without large throughput penalties in our system ($<$2\% loss), any decentration between the fiber and lens can cause significant light loss (3\% loss for 10$\upmu$m fiber-core offset, 8\% loss for a $\sim$20 $\upmu$m offset.) This drove our concentricity tolerance requirements for both the v-groove and FC connector prototype devices. Rotational misalignment of the two octagonal faces of connecting fibers can cause up 4\% additional loss (2\% on average is carried for our efficiency estimation). Note that alignment tolerances are dictated by the diameters of the fiber and ball lens, and must be calculated individually for each instrument.

\begin{figure}
\begin{center}
\includegraphics[width=3.4in]{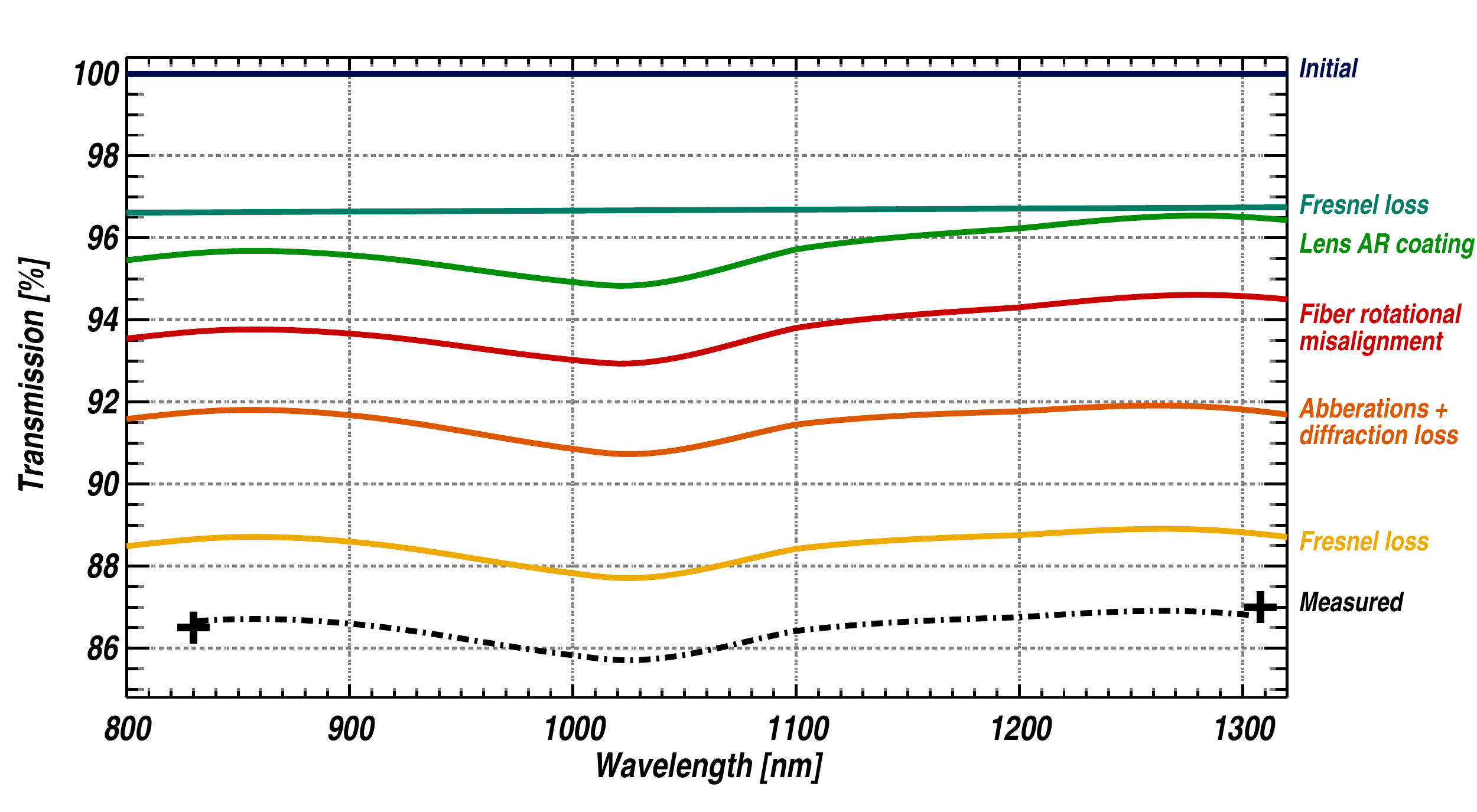}
\caption{Theoretical throughput budget of double scrambler system with a BBAR coating on the ball lens and two (uncoated) 300 $\upmu$m octagonal fibers. Our measured throughput values at 830 and 1310 nm (black crosses) are close to the theoretical maximum for this configuration. The dashed line is the theoretical throughput curve shifted to the measurement values.}
\label{fig:throughput}
\end{center}
\end{figure}

With uncoated fibers, a BBAR coating on the ball, and a simple alignment scheme, we measure 85-87\% throughput for both prototype devices using 830 nm and 1310 nm fiber lasers. The predicted maximum for this configuration is $\sim$89\% at these wavelengths. Figure~\ref{fig:throughput} shows the theoretical throughput for the prototype device in its current configuration. 

The stainless steel connectors used in our FC connector prototype have typical tolerances of 5 $\upmu$m between the center of the connector bore to the outer diameter of the ferrule. This prototype has the benefit of practical rotational alignment ability, as the fibers can be freely rotated inside the mating sleeve to better match octagonal core geometries. The v-groove mount has somewhat better concentricity due to the better manufacturing tolerance ($<$3~$\upmu$m vs 5~$\upmu$m), but does not easily allow for repeatable rotation of the fiber cores. Our measured throughput for the sleeve and connector prototype could likely be improved by using custom, tight tolerance ($<$0.5 $\upmu$m) ferrules rather than standard commercial connectors to maintain concentricity between the lens and the fiber centers.

Our theoretical maximum design throughput, with dual-band AR or broadband AR (BBAR) coatings on both the ball and the fiber faces, and perfect alignment of the octagonal cores, is roughly 94\% for our wavelength range. The inefficiency of previous double scrambler designs was the principal deterrent in their common use. The high efficiency of this design now enables the achievement of very high scrambling gains with a minimal overall throughput penalty.

\begin{center}
\begin{deluxetable}{ccc}
\tablewidth{0pt}
\tablecaption{Efficiencies and laboratory scrambling gains of optical fiber double-scrambler systems.}
\tablehead{
\colhead{Reference}  & \colhead{Efficiency}	& \colhead{SG (if reported)}
}
\startdata
\cite{Hunter:1992}	&	20\%		&	-	\\
\cite{Casse:1997}	&	20\%		&	-\\
\cite{Avila:1998}	&	68\%		&	-\\
\cite{Raskin:2008}	&	70\%	&		1100\\
\cite{Barnes:2010}	&	75\%		&	-\\
\cite{Avila:2012}	&	70\%		&	6000\\
\cite{Chazelas:2012}	&	$\sim$80\%	&	$\sim$10,000	\\
\cite{Spronck:2013}	&	65\%		&	-\\
This work\footnote{Measured using 830 \& 1310 nm fiber lasers} (oct + DS + oct)		&	85\%\footnote{FC connectors with ceramic mating sleeve}, 87\%\footnote{V-groove mount}		& 	13,000	

\enddata
\label{tab:eff}
\end{deluxetable}
\end{center}

\section{Summary}
The ideal scrambler for fiber fed instruments is effective at scrambling both the fiber near and far fields, while still being easy to align and efficient. The high-index ball-lens scrambler, when used with a combination of octagonal fibers, satisfies all of these quantities in a single, compact system. This device is being developed for the near-infrared Habitable-zone Planet Finder spectrograph on the 10-m Hobby Eberly Telescope, but its versatility implies that it could easily be adapted for other current and future fiber-fed precision radial velocity spectrographs such as the CARMENES \citep{Quirrenbach:2014}, MINERVA \citep{swift:2014}, NRES \citep{eastman:2014}, and PARAS \citep{Chakraborty:2008} spectrographs. Since the HET has a uniquely variable illumination pupil, the HPF fiber train will require a configuration that maximizes scrambling, i.e. a combination of octagonal fibers, circular fibers, and a double-scrambler to minimize the scrambling-induced velocity errors. 

We have demonstrated the high scrambling performance of this prototype double-scrambler system by examining the extreme cases of input illumination variation. Pairing the ball lens scrambler with octagonal fibers yields scrambling gains in excess of 10,000 with high thoughput (87\%). Adding a circular fiber further improves the scrambling gain to $>$20,000, limited by our measurement apparatus. This translates to roughly 10~cm~s$^{-1}$ velocity error for the HPF instrument due to telescope guiding error, exceeding our goal of 30~cm~s$^{-1}$ guiding-induced velocity measurement error.

In replicating this system for other fiber-fed spectrographs it might not be necessary to include every scrambling element, but the inclusion of a compact efficient ball-lens scrambler would stabilize instrument illumination and could substantially benefit many high precision spectrographs. Decoupling the instrument PSF from the telescope is a major technical challenge that will need to be overcome for the next generation of extreme precision planet-hunting spectrographs that aim to detect Earth-mass extra-solar planets with precisions of $\sim$10 c\ms \citep{pepe:2014}.

Due to the monolithic nature and simplistic design, this ball lens scrambler could be easily replicated on larger scales for multi-object instruments as well, aiding in precision spectroscopy for upcoming galactic and extra-galactic surveys.

\acknowledgments{S. Halverson and A. Roy contributed equally to this work. We thank the anonymous referee for helpful suggestions and comments. This work was partially supported by funding from the Center for Exoplanets and Habitable Worlds. The Center for Exoplanets and Habitable Worlds is supported by the Pennsylvania State University, the Eberly College of Science, and the Pennsylvania Space Grant Consortium. We thank Polymicro for sample reels of octagonal fiber. We acknowledge support from NSF grants AST 1006676, AST 1126413, AST 1310885, and the NASA Astrobiology Institute (NNA09DA76A) in our pursuit of precision radial velocities in the NIR. SPH and AR acknowledge support from the Penn State Bunton-Waller, and Braddock/Roberts fellowship programs and the Sigma Xi Grant-in-Aid program. }
\newpage

\bibliographystyle{apj}
%\bibliography{ms}

\end{document}